\documentclass[5p,times,twocolumn]{elsarticle}

\usepackage{amsmath}

\DeclareMathOperator*{\argmin}{arg\,min}
\usepackage{pifont}
\newcommand{\cmark}{\ding{51}}%
\newcommand{\xmark}{\ding{55}}%
\usepackage{bm}

\usepackage{algpseudocode}
\usepackage[linesnumbered,ruled,vlined]{algorithm2e}
\SetKwInOut{KwIn}{Input}
\SetKwInOut{KwOut}{Output}

\usepackage{enumitem}
\usepackage{graphicx}
\usepackage{subcaption}
\newcommand{\tabincell}[2]{\begin{tabular}{@{}#1@{}}#2\end{tabular}}
\usepackage{mathrsfs}
\usepackage{color}
\usepackage{lineno,hyperref}

\usepackage{multirow}

\journal{Journal of \LaTeX\ Templates}

\begin{document}

\begin{frontmatter}

\title{Toward multi-target self-organizing pursuit in a partially observable Markov game\tnoteref{mytitlenote}}
\tnotetext[mytitlenote]{
This work is partially supported by the Shenzhen Fundamental Research Program under Grant No. JCYJ20200109141235597, the National Science Foundation of China under Grant No. 61761136008, the Shenzhen Peacock Plan under Grant No. KQTD2016112514355531, the Program for Guangdong Introducing Innovative and Entrepreneurial Teams under Grant No. 2017ZT07X386, and  the Australian Research Council (ARC) under Discovery Grant DP210101093 and DP220100803.
}

%
%


\author[sustechaddress,utsaddress]{Lijun Sun}
\ead{Lijun.Sun@student.uts.edu.au}

\author[utsaddress]{Yu-Cheng Chang}
\ead{Yu-Cheng.Chang@uts.edu.au}

\author[swuaddress]{Chao Lyu}
\ead{lyuchao@swu.edu.cn}

\author[sistaddress]{Ye Shi}
\ead{shiye@shanghaitech.edu.cn}

\author[sustechaddress]{Yuhui Shi\corref{mycorrespondingauthor}}
\cortext[mycorrespondingauthor]{Corresponding author: Yuhui Shi, Chin-Teng Lin.}
\ead{shiyh@sustech.edu.cn}

\author[utsaddress]{Chin-Teng Lin\corref{mycorrespondingauthor}}
\ead{Chin-Teng.Lin@uts.edu.au}

\address[sustechaddress]{Guangdong Provincial Key Laboratory of Brain-inspired Intelligent Computation, Department of Computer Science and Engineering, Southern University of Science and Technology, China}
\address[utsaddress]{Centre for Artificial Intelligence, CIBCI Lab, Faculty of Engineering and Information Technology, University of Technology Sydney, Australia}
\address[swuaddress]{College of Computer and Information Science, Southwest University, China}
\address[sistaddress]{School of Information Science and Technology, ShanghaiTech University, China}

\begin{abstract}
The multiple-target self-organizing pursuit (SOP) problem has wide applications and has been considered a challenging self-organization game for distributed systems, in which intelligent agents cooperatively pursue multiple dynamic targets with partial observations.  
This work proposes a framework for decentralized multi-agent systems to improve the implicit coordination capabilities in search and pursuit.
We model a self-organizing system as a partially observable Markov game (POMG) featured by large-scale, decentralization, partial observation, and noncommunication. 
The proposed distributed algorithm–fuzzy self-organizing cooperative coevolution (FSC2) is then leveraged to resolve the three challenges in multi-target SOP: distributed self-organizing search (SOS), distributed task allocation, and distributed single-target pursuit. 
FSC2 includes a coordinated multi-agent deep reinforcement learning (MARL) method that enables homogeneous agents to learn natural SOS patterns. 
Additionally, we propose a fuzzy-based distributed task allocation method, which locally decomposes multi-target SOP into several single-target pursuit problems. 
The cooperative coevolution principle is employed to coordinate distributed pursuers for each single-target pursuit problem. 
Therefore, the uncertainties of inherent partial observation and distributed decision-making in the POMG can be alleviated. 
The experimental results demonstrate that by decomposing the SOP task, FSC2 achieves superior performance compared with other implicit coordination policies fully trained by general MARL algorithms.
The scalability of FSC2 is proved that up to 2048 FSC2 agents perform efficient multi-target SOP with almost 100\% capture rates.
Empirical analyses and ablation studies verify the interpretability, rationality, and effectiveness of component algorithms in FSC2.
\end{abstract}


\end{frontmatter}


\section{Introduction}\label{sec_intro}

\textit{\textbf{Self-organizing systems and multi-agent coordination without communication.}}
Self-organization is a type of swarm intelligence that can be found in natural environments and animal behaviors: rippled sand dunes, synchronized flashing fireflies, fish schooling, flocking birds, etc \cite{camazine2001selforganization}.
It forms order and structure through purely internal and local interactions in a system, without any external controls.
So, many researchers \cite{rubenstein2014programmable, Berlinger2021Implicit, warnat2021swarmlearning, ye2016survey} made efforts to understand the nature and create artificial self-organization systems that can be characterized by decentralization, partial observation, scalability, and emergent properties.
However, in general multi-agent game setups, communication failures cannot be avoided due to communication attacks, varying protocols, blocked channels, physical distance, damage, energy conservation, etc.
In such scenarios, multi-agent coordination degrades, since no commands, role assignments,  conflicts elimination, information sharing, or other negotiations can be exchanged among agents.
Therefore, more effective implicit coordination is expected for the more restricted self-organizing setup that does not rely on communication.

\textit{\textbf{Background and application of the pursuit problem.}} 
This work investigates the multiple-target self-organizing pursuit (SOP) problem.
It formulates general competitive and cooperative interactions among agents and thus can serve as a basic capability of agents in standardized problems and real-world applications. 
In warfare, agents may be any confrontational devices, such as fighters, bombers, and missiles \cite{isaacs1965differential}.
In aerospace, one goal is to clean up space debris, inactive satellites, and military vehicles to ensure the safety of active space assets or aerial vehicles \cite{Aerospace2020Satellite, Guan2021Bounded, aerospace2021Incomplete, Aerospace2021PursuitEvasion}.
The searchers, pursuers, targets, or evaders in the pursuit may also represent 
players in a football game,
lions and humans in a bounded arena \cite{littlewood1953mathematician}, 
searchers and lost spelunkers in a cave \cite{Parsons1978},  
cops and robbers in a city \cite{NOWAKOWSKI1983235,fomin2008tractability,bonato2011game},
pollutants and cleaning robots in the environment \cite{chung2011search}, 
creatures in biological systems \cite{camazine2001selforganization}, etc.

\textit{\textbf{Self-organizing pursuit game setup in comparison with representative MARL pursuits.}}
Comparing existing popular pursuit environments, MPE (multi-agent particle environments) pursuit \cite{lowe2017multi} uses the occupying-based capture, where one pursuer occupies the same position of a target or their distance is smaller than a threshold.
However, its global observation representation scales poorly with the number of agents.
MAgent pursuit (battle) \cite{Zheng2018MAgent} uses the tag-based capture, where a target is attacked if it is tagged by pursuers.
Besides, it provides the option to use the global information in the observation or not.
The mean field pursuit \cite{ZHOU2021} solves one mass capture problem, where the mass center of the pursuers matches that of the targets is called a capture.
Although it can be applied in the large-scale pursuit, the one mass capture is totally different from the multi-target captures that are distributed in the whole space.
Moreover, none of these environments consider the interagent collision avoidance problem.

Compared with the above capture definitions, the surrounding-based capture in most literatures is more general in terms of multi-agent behaviors and challenging in terms of coordination, where a target cannot move only if it is surrounded by pursuers.
Therefore, this work builds the multi-target self-organizing pursuit (SOP) environment and consider a more practical and challenging multi-agent setting: large-scale partially observable pursuers coordinate without explicit communications, search, chase, and surround multiple distributed dynamic targets until all targets are found and captured without collisions in a grid world.
It is worth noting that we consider and report the multi-agent collision avoidance performance, which is a crucial metric concerning the safety in deploying the multi-agent system (MAS) but is rarely reported especially in general MARL literatures.

\textit{\textbf{Related work and MARL coordination solution.}}
Since Isaacs \cite{isaacs1965differential}, the differential games are used to formulate the pursuit problem, which look for saddle-point strategies and model the dynamics in games with differential equations \cite{Aerospace2020Satellite, Guan2021Bounded, aerospace2021Incomplete, Aerospace2021PursuitEvasion, Weintraub2020Introduction}.
However, most such works relate to two-agent zero-sum games.
Another seminal work proposed by Benda et al. \cite{benda1986optimal} explores the pursuit problem to investigate the optimal communication structure of agents.
In addition to the conventional communication with predefined communication topology, message content, and transmission frequency, selective communications \cite{wang2020cooperative, hejazi2021multi, xiao2023graph}, including dynamic event-triggered communications \cite{lv2022non,wang2022distributed}, are studied.
When there are no communications, agents have no ways to get a bigger picture of the world by actively exchanging information. 
Therefore, it is more challenging to make decision only with agents' own partial observations and the uncertain behaviors of other agents, i.e., the interaction uncertainty \cite{Mykel2022DecisionMaking}.
Finally, although many previous works consider the surrounding-based capture, they model the pursuit domain with the Markov decision process (MDP), where each agent can fully observe all agents' positions.

On the other hand, in terms of  the partially observable multi-agent settings, many general multi-agent reinforcement learning (MARL) algorithms are tested in the pursuit domain.
Coordinated agents can outperform fully independent agents \cite{tan1993multi}.
In particular, the centralized training and decentralized execution (CTDE) is a general framework of coordinated learning for decentralized multi-agent systems.
One effective way of implementing CTDE is to apply the concept of parameter sharing \cite{Gupta2017}, which enables the extension of single-agent reinforcement learning (RL) algorithms to the multi-agent setting, such as the actor-critic algorithm \cite{sutton2018reinforcement}.
It is extremely useful for the learning of large-scale homogeneous agents by training shared (value or policy) models from collective experiences.
Besides, it also benefits the coordinated learning efficiency as shown in  the experiments of \cite{terry2020revisiting,yu2022surprising}.

To enhance the cooperation of agents, many CTDE MARL algorithms use the centralized (action) value functions, such as MADDPG \cite{lowe2017multi}, COMA \cite{foerster2018counterfactual}, QMIX \cite{rashid2018qmix}, and MAPPO \cite{yu2022surprising}.
A common issue of such centralized (action) value function learning is that the computational complexity increases with the number of agents involved and the trained agents are heterogeneous, which hinder the large-scale deployment.
Besides, since these centralized (action) value functions are optimized with a fixed number of agents, when the agent swarm size changes, the policy is hard to guarantee its optimality and new training is needed.
For example, in the MADDPG, each agent separately maintains a centralized critic function that takes the joint observation and joint action as inputs, while agents' actor functions use only local information.
In contrast, although MAPPO also learns a centralized value function that accesses the global information out of the partial observations of the agents, it uses the parameter sharing for homogeneous agents and thus potential for large-scale applications.
Mean field reinforcement learning \cite{yang2018mean} also uses the CTDE framework and tackles large scale multi-agent problems by simplifying the interactions between agents to the interplay between an agent and the mean effect of its neighborhood, i.e., the virtual mean agent, through the mean field approximation.
However, by employing the mean field theory, it explicitly ignores the detailed interactions between real agents, which means it cannot deal with the collision avoidance between agents, i.e., the safety RL issues.
Another way to enhance the cooperation is to allow communications.
For example, DGN (graph convolutional reinforcement learning) \cite{jiang2019graph} achieves the cooperation of agents through local communications to interchange the intermediate outputs of agent models.
As for other MARL works tested regarding pursuit, they are mostly  subject to several or all of the constraints: small-scale, fully observable pursuers, single-target pursuit, occupying-based capture, with communications, and permitted collisions (see also \cite{Souza2021Decentralized}).

\begin{figure*}[htbp!]
	\centering
	\includegraphics[width=0.7\linewidth]{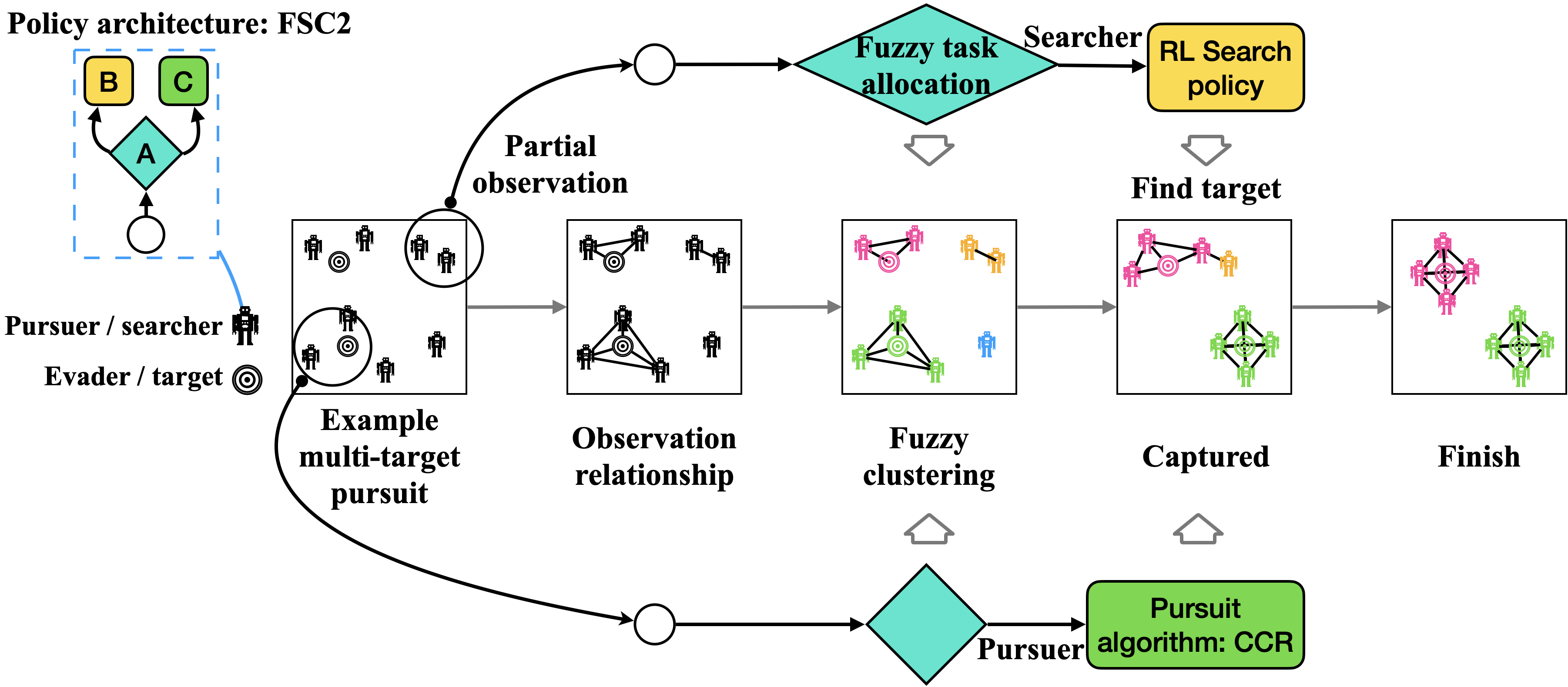}
	\caption{\textbf{Multi-target self-organizing pursuit task and computational framework of FSC2} (fuzzy self-organizing cooperative coevolution). 
FSC2 is a distributed framework that consists of three modules: 
\textbf{A. fuzzy task allocation:} takes partial observation as input, computes distributed fuzzy clusters of agents and targets, and determines the current role of agent to be either a searcher or a pursuer; 
\textbf{B. RL search policy:} a searcher searches the space to find targets; 
and \textbf{C. pursuit algorithm–CCR:} a pursuer cooperates with cluster members to pursue the targeted evader of its cluster. 
}
	\label{fig_framework_fsc2}
\end{figure*}

\textit{\textbf{Our work and contribution.}}
Based on the above discussions, the main contributions of this paper are:
\begin{itemize}
\item To facilitate the study of implicit multi-agent coordination without communications,
this work fills the current literature gap in the self-organizing pursuit (SOP) setup featured by large-scale, decentralization, partial observation, no communication, no interagent collision, multiple distributed targets, and surrounding-based capture.
To enable this study, we have built the SOP environment \footnote{All code is available at https://github.com/LijunSun90/pursuitFSC2.}.
\item To address the severer interaction uncertainty \cite{Mykel2022DecisionMaking} and observation uncertainty due to no communications, this work proposes the distributed hierarchical framework called the fuzzy self-organizing cooperative coevolution (FSC2) for the multi-target SOP, as shown in Figure \ref{fig_framework_fsc2}.
Through analysis, it decomposes the SOP into three sub-problems that can be well formulated and can thus utilize the strengths of fuzzy logic, MARL, and evolutionary computation (EC).
\end{itemize}

Further, the innovations of the proposed FSC2 framework can be summarized as follows.
\begin{itemize}
\item 
The first module of FSC2: fuzzy based task allocation overcomes the consensus issue of the distributed clustering in two folds by the fuzzy logic with introduced memory.
First, to improve the consensus between independent agents without communications, we utilize the fuzziness of fuzzy clustering in identifying the cluster memberships of agents. 
Second, to keep a consistent clustering decision of a single agent in the time scale, we introduce an incremental agent memory in the distributed fuzzy clustering.
\item 
The second module of FSC2 for search learns reasonable and explainable behaviors for large-scale homogeneous agents by the CTDE actor-critic algorithm. 
By formulating as the partially observable Markov game (POMG) and designing the reward function, the unknown self-organization mechanism can be learned that maps the local search policy to the coordinated global space exploration without communications.
\item 
The third module of FSC2: pursuit algorithm–CCR proposes a distributed coordination mechanism that can ensure the safety of multi-agent collision avoidance in the target pursuit within clusters.
To conquer the partial observation uncertainty and the limit of no communications, the coevolutionary coevolution scheme is used for the online planning in balancing the individual and swarm interests, while the lexicographic convention is adopted for close coordination with the introduced concept of certain partial observation.
\end{itemize}

The organization of the paper is as follows.
First, the problem formulation of self-organizing pursuit is given in Section \ref{sec_problem_formulation}.
Second, the proposed approaches are given in Section \ref{sec_proposed}.
Third, the experimental results, analyses, and discussions are given in Section \ref{sec_experiments}.
Finally, the conclusions, limitations, and future work are given in Section \ref{sec_conclusion}.

%
%

\section{Problem formulation}\label{sec_problem_formulation}

\subsection{Multi-agent formulation of self-organization systems}

\begin{figure}[htbp!]
	\centering
	\subfloat[Common multi-agent problem formulations \cite{Mykel2022DecisionMaking}.]{\includegraphics[width=1.\linewidth]{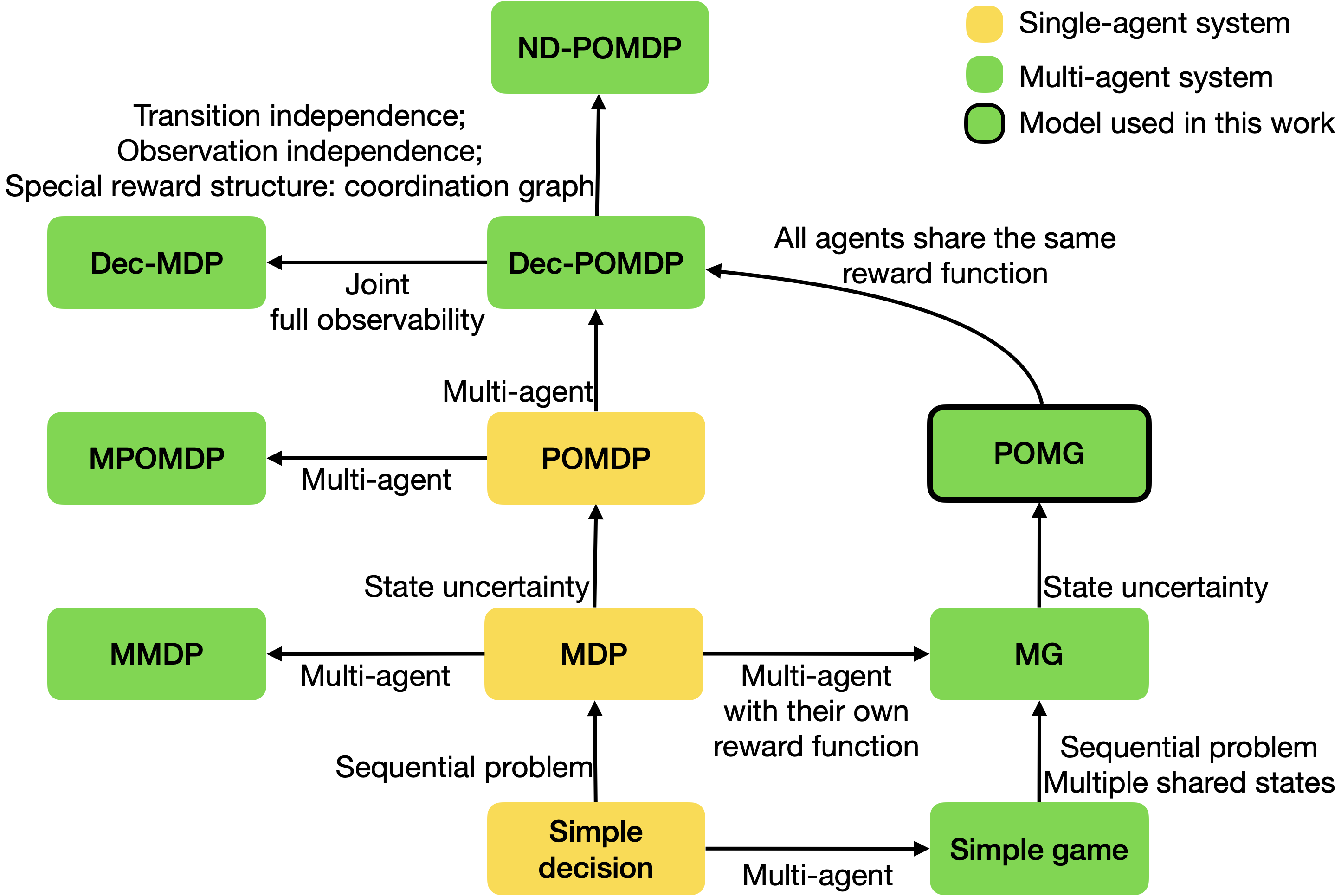}
	\label{fig_mas_problem_formulation}}
	\\
	\subfloat[Common nomenclature for multi-agent models.]{\makebox[1.2\width]{\includegraphics[width=0.7\linewidth]{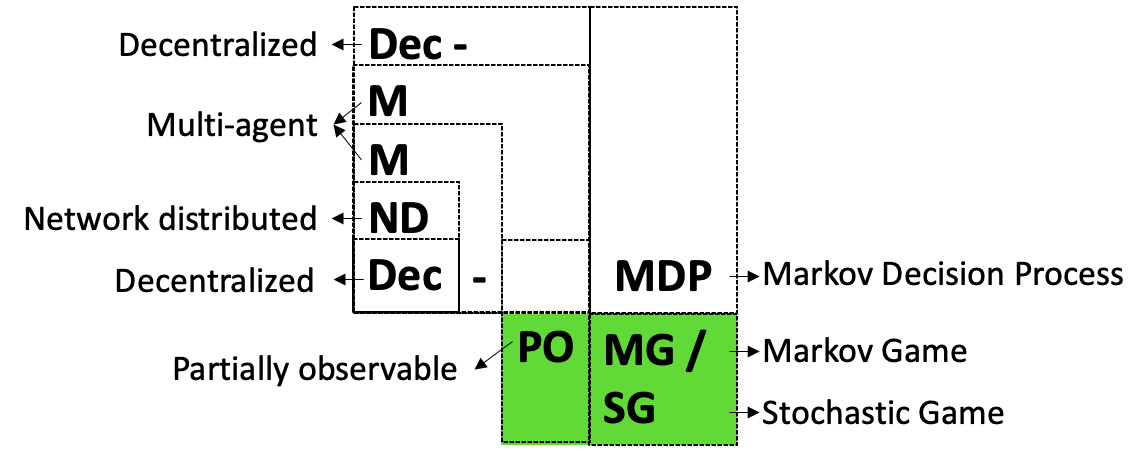}}
	\label{fig_mas_model_names}}	
	\caption{Common multi-agent problem formulations and their nomenclature.}
	\label{fig_capture_state}
\end{figure}

A multi-agent system (MAS) can be seen as a decision-making system in  which each agent is a decision maker.
It can be formulated in terms of the following four factors: 
(1) the number of agents: a single agent or multiple agents;
(2) state transitions: present (sequential problem) or not;
(3) the uncertainty of observability: full observability, joint full observability, or partial observability; and 
(4) the reward function: each agent has an individual reward function, all agents share the same reward function, or different groups of agents have separate reward functions.
Based on the above four dimensions, various models have been proposed and investigated, as shown in Figure \ref{fig_mas_problem_formulation} \cite{Mykel2022DecisionMaking}, 
and the common nomenclature for the model name abbreviations is presented in Figure \ref{fig_mas_model_names}.

A definition of self-organization was given in \cite{camazine2001selforganization}: global level patterns unexpectedly emerge solely from the distributed decentralized local nonlinear interactions of components of the system under behavioral rules (of thumb) with local information and no external directing influences.
In terms of these features, we can formulate a self-organizing system as a POMG \cite{Mykel2022DecisionMaking}: 
$\langle \gamma, \mathcal{I}, \mathcal{S}, \mathcal{A}, \mathcal{O}, P, O, R \rangle$. 
$\gamma$ is the discounted factor for return; 
$\mathcal{I}=\{1, ..., n\}$ represents all total $n$ agents; 
$\mathcal{S}=\{s\}$ is the true state space; 
$\mathcal{A}=\mathcal{A}^1 \times ... \times \mathcal{A}^n = \{\vec{a}\}$ is the joint action space; 
$\mathcal{O}=\mathcal{O}^1 \times ... \times \mathcal{O}^n = \{\vec{o}\}$ is the joint observation space;
$P(s'|s, \vec{a})$ is the transition function from the current state $s$ to the next state $s'$ given the joint action $\vec{a}$;
$O(s)=\{o^1(s),...,o^n(s)\}$ is the joint observation function; 
and $R(s, \vec{a})=\{R^1(s, \vec{a}),...,R^n(s, \vec{a})\}$ is the joint reward function and each agent maximizes its own accumulated reward.

The reason we use the POMG rather than the Dec-POMDP (decentralized partially observable Markov decision process) to represent a self-organizing system is that in the Dec-POMDP, all agents are fully cooperative in that they aim to maximize a collective reward $R(s, \vec{a})$, while in a general self-organizing system, even collaborative agents have unequal rewards and need to balance the swarm benefits and their own benefits. 
Therefore, POMG is more similar to the natural swarm intelligence.

\subsection{The problem of self-organizing search and pursuit}

\begin{figure}[htb!]
	\centering
	\includegraphics[width=0.45\linewidth]{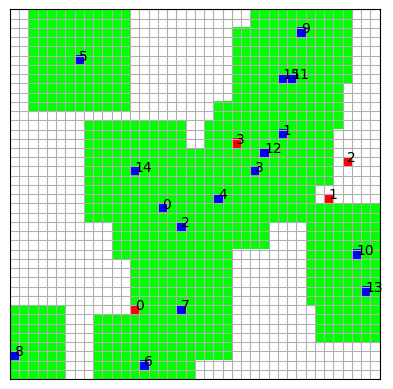}
	\caption{A screenshot of self-organizing search and pursuit in a bounded grid world, where {\color{red}red} squares are targets or evaders, {\color{blue}blue} squares are searchers or pursuers, and {\color{green}green} background around each agent shows its perception range with an $inf$-norm radius of 5.}
	\label{fig_multiple_preys_pursuit}
\end{figure}

A typical multi-target search and pursuit scenario is illustrated in Figure \ref{fig_multiple_preys_pursuit}.
Due to the partial observation and communication limitation of agents, we distinguish the self-organizing search (SOS) and self-organizing pursuit (SOP) as two different but related multi-agent problems, where the search policy in SOS is taken as a basic capability of agents in the SOP.
\begin{itemize}[leftmargin=*]
\item \textbf{Self-organizing search (SOS):} A search is considered successful when a searcher occupies the same position of a target, and the target will then disappear.
The SOS terminates when all targets in the environment disappear or the maximum time is reached.
\item \textbf{Self-organizing pursuit (SOP):} A capture is considered as successful when a target is encircled by four pursuers and cannot move further.
However, the target will not disappear after it is captured in the SOP.
The game terminates when all targets are found and captured or the maximum time is reached.
\end{itemize}

Note that, the SOS task is only used to train the search policy in Figure \ref{fig_framework_fsc2}, i.e., the space exploring ability that will be used in the SOP task.
We design the SOS task harder than the search requirement in the SOP to better train the search policy.
First, the SOS task uses multiple static targets since searching for static targets are sometimes harder than dynamic ones in bounded environments, as the agent has no chance to wait for the target coming.
In addition, in the SOS task, a target is designed to disappear after being searched to make the search task harder and harder with time, especially when there is no communication and information exchange between agents.
Last, in the actual pursuit, agents are not expected to collide with the targets or evaders.
However, in the SOS task, we specially define a successful search as that a searcher occupies a target rather than a target appears in the agent's local view, which also only serves the purpose of training.
This is because, in the pursuit where the search policy is applied, more than one agents are expected to find and approach the same target simultaneously in order to finally capture it.

In the following, we investigate the coordination strategies for agents constrained by: 
(1) the observation range of an agent is the scope of radius 5 according to the $inf$-norm, i.e., an $11\times11$ square centered at the agent;
(2) communication between agents is limited that they can only see the positions of targets and other agents in their own local views, and no other information exchange is allowed;
(3) the available movements of all agents are 5 discrete actions \{up, down, right, left, still\} in the grid world. 
Therefore, the $inf$-norm is used in the agent's perception, and the 1-norm (Manhattan distance) is used in the agent's movement, which are widely adopted in MAS.

\section{Proposed approach for self-organizing pursuit (SOP)}\label{sec_proposed}	

In this section, we introduce in detail the proposed distributed hierarchical framework–fuzzy self-organizing cooperative coevolution (FSC2) in Figure \ref{fig_framework_fsc2}.
FSC2 is a distributed algorithm for homogeneous swarm of agents that each agent consists of three modules: (1) fuzzy clustering; (2) search policy; and (3) pursuit algorithm–CCR.
Its main idea and motivation is to decompose the distributed self-organizing pursuit (SOP) problem into sub-tasks that are more intuitive  and simpler to be well defined and solved.

The whole algorithm of FSC2 is given in Algorithm \ref{algorithm_FSC2}.
In the multi-target pursuit environment, targets and partial observable agents are distributed in the space.
First, we assume two alternate basic roles of an agent: searcher or pursuer, based on the existence of free targets that are not captured in the agent's neighborhood.
Then, agents are distributed clustered that each searcher forms a separate cluster and pursuers are clustered based on their neighborhood relationships.
This clustering process is conducted by the first module–fuzzy clustering algorithm in Section \ref{sec_fuzzy_clustering}, and Figure \ref{fig_framework_fsc2} gives an illustrative clustering result.
After clustering, an agent alternates between the second module: search policy in Section \ref{sec_sos} and the third module: pursuit algorithm–CCR in Section \ref{sec_ccr}, based on its real-time neighborhood.


\begin{algorithm}[htb!]
\caption{FSC2 for each agent in the SOP}
\label{algorithm_FSC2}
\While{the termination conditions are not satisfied}{
	$role, cluster\_center, cluster\_members, Memory \gets$ Fuzzy clustering (Algorithm \ref{algorithm_fuzzy_clustering} in Section \ref{sec_fuzzy_clustering}). \\
	\If{$role$ is a searcher}{
		As an SOS agent (Section \ref{sec_sos}), find free targets.
	}
	\ElseIf{$role$ is a pursuer}{
		As a CCR agent (Section \ref{sec_ccr}), cooperate with $cluster\_members$ in pursuing $cluster\_center$.
	}
}
\end{algorithm}

\subsection{Distributed fuzzy clustering for task allocation}\label{sec_fuzzy_clustering}
We define that a pursuer is free if it has not captured a target, while a target is free if it has not been captured.
So, an agent is either a searcher, which explores the space to find a free target, or a pursuer, which cooperates with other free pursuers to capture a free target.
In the multi-target SOP, since four pursuers are required to capture each target, distributed task allocation or clustering is needed to determine which group of pursuers capture which free target.

The main challenge in the multi-agent distributed clustering is the consensus issue in two folds due to 
the partial observation uncertainty and the interaction uncertainty.
First, since agents cannot fully observe the world or share the same knowledge through communications, they cannot independently make exactly the same decision.
To address this issue, we adopt the fuzzy clustering and utilize its fuzziness in identifying the cluster memberships to reach a consensus with a higher probability.
Second, an agent may frequently switch between the roles of searcher and pursuer over a short period of time steps due to its partial observability, which causes instability in the distributed clustering. 
We, therefore, introduce an incremental agent memory in the fuzzy clustering.

\textit{\textbf{Fuzzy membership.}}
Since the task of the pursuers is to capture targets, for agent $k$, the cluster centers are all its $m^k$ local free targets $T = \{T_1, ..., T_{m^k} \}$, while the $n^k$ local free pursuers $A = \{A_1, ..., A_{n^k} \}$ need to be clustered, and both $T_j$ and $A_i$ are 2-D positions.
The fuzzy membership value of the free pursuer $A_i$ with respect to the cluster center $T_j$ in agent $k$'s view is calculated by
\begin{equation}\label{eq_mu_ij}
\mu^k_{ij} = \frac{(||A_i - T_j||_1^2)^{\frac{1}{1-\alpha}}}{\sum_{j=1}^{m^k}(||A_i - T_j||_1^2)^{\frac{1}{1-\alpha}}} \in [0, 1],
\end{equation}
where $\alpha > 1$ is the fuzzifier \cite{bezdek2013pattern}, the value of which is 1.5 in our experiments.
Thus, agent $k$ can obtain its fuzzy membership matrix 
\begin{equation}\label{eq_fuzzy_membership_matrix}
M^k = [\mu^k_{ij}] \in R^{n^k \times m^k},
\end{equation}
the $i$-th row $M^k_{i*}$ of which is the fuzzy membership value of agent $i$ with respect to all local cluster centers in agent $k$'s point of view. 
Based on $M^k$, agent $k$ can obtain its membership matrix 
\begin{equation}\label{eq_membership_matrix}
\hat{M}^k \sim M^k,
\end{equation}
which is a binary matrix.
Its only one element with the value 1 in the $i$-th row $\hat{M}^k_{i*}$ is sampled from the random distribution determined by $M^k_{i*}$, since an agent can only belong to one cluster.
Based on $\hat{M}^k$, the cluster center of agent $k$ is the target 
\begin{equation}\label{eq_cluster_center}
T_c |_{\hat{M}^k_{kc} \ne 0, c = 1, ..., m^k},
\end{equation}
while agent $k$'s cluster members are the pursuers 
\begin{equation}\label{eq_cluster_members}
\{A_i | \hat{M}^k_{ic} \ne 0, i = 1, ..., n^k \}.
\end{equation}
The distributed fuzzy clustering based task allocation process in Equation (\ref{eq_mu_ij}) to (\ref{eq_cluster_members}) is summarized in Algorithm \ref{algorithm_fuzzy_clustering}.

\textit{\textbf{Agent memory.}}
Note that, each agent's $Memory$ of the environment (line 1 of Algorithm \ref{algorithm_fuzzy_clustering}) is updated through its experiences, which includes the captured status of targets and locked status of pursuers.
So, the maximum size of $Memory$ is the same for all pursuers, which is determined by the possible number of targets and pursuers in the environment.
Without a $Memory$, an agent may oscillate between the roles of a searcher and a pursuer.
For instance, an agent may walk one step closer to a target, see the target captured by 4 pursuers, and know that itself is a searcher; if it then walks one step away from the target, the agent can only see 3 pursuers surrounding the target and cannot identify for certain whether it is captured, although it previously observed its captured status.
In other cases, a target may be falsely captured such as when it is only blocked by another free target.
When that free target walks out of its way, the previous ``captured" target becomes free again.
In such scenarios, the agent should also update its $Memory$ when it is pretty sure based on its newest observation.

In addition, note that, although the number of local clusters is determined by the number of local free targets in Equation (\ref{eq_mu_ij}), the number of members in each cluster is not specified in Equation (\ref{eq_cluster_members}).
So, it is possible that more pursuers are clustered into one same nearer target while less pursuers to a farther one.
It may be a bit greedy and redundant sometimes that pursuers first cooperate to capture one nearer target as soon as possible and then pursue others.
However, this redundancy in the self-organizing clustering may improve the system's robustness to individual robot's software or hardware failures.
%
\begin{algorithm}[htb!]
\caption{Distributed fuzzy clustering of agent $k$}
\label{algorithm_fuzzy_clustering}
\KwIn{local observation $o^k_t$ of agent $k$ at time $t$.} 
\KwOut{$role, cluster\_center, cluster\_members, Memory$.} 
Update captured targets and locked pursuers in $Memory$. \\
\If{there are no local free or neighboring targets}{
	$role \gets$ searcher.\\
	$cluster\_center \gets$ the agent itself $A_k$.\\
	$cluster\_members \gets$ the agent itself $A_k$.
}
\Else{
	$role \gets$ pursuer.\\
	$T = \{T_1, ..., T_{m^k} \} \gets$ local free targets.\\
	$A = \{A_1, ..., A_{n^k} \} \gets$ local free pursuers. \\
	$cluster\_center \gets$ Equation (\ref{eq_cluster_center}).\\
	$cluster\_members \gets$ Equation (\ref{eq_cluster_members}).
}
\end{algorithm}

\textit{\textbf{Global distributed consistency metric.}}
To evaluate the consistency in the distributed clustering process between the global $n$ agents and $m$ targets, a consistency matrix $C = [c_{ij}] \in R^{n \times n}$ can be calculated from $\{\hat{M}^k | k = 1, ..., n \}$.
$c_{ij} \in \{-1, 1, ..., m\}$ is the global target index of the non-zero item of $\hat{M}_{j*}^i$, which represents the cluster (or target) index for agent $j$ from agent $i$'s point of view, and $c_{ij}=-1$ means that agent $i$ has no idea of the cluster of agent $j$ because agent $j$ is located out of the local view of agent $i$.

The global DC (distributed consistency) can be defined as
\begin{equation}\label{eq_dc}
\begin{split}
DC & \doteq 
\frac{2}{n \cdot (n-1)} 
\sum_{i=1}^{n-1} \sum_{j=i}^{n} 
\frac{|\{k | k \in \hat{C_i} \cap \hat{C_j}, \text{and } c_{ik} == c_{jk}\}|}
{|\hat{C_i} \cap \hat{C_j}|} \\
& \in [0, 1],
\end{split}
\end{equation} 
where $|\cdot|$ is the the cardinality of a set; 
$\hat{C_i} = \{k | k = 1, ..., n, \text{and } c_{ik} \ne -1 \}$ is the set of visible local pursuers for agent $i$.
The process of computating $DC$ in Equation (\ref{eq_dc}) is to compare every two rows $C_{i*}$ and $C_{j*}$ of $C$ and calculate the ratio of consistent decisions between agent $i$ and agent $j$ in their common knowledge about the other pursuers.
Due to this special meaning in our application, we define $0/0 = 1$ for Equation (\ref{eq_dc}), which means that two agents without local physical interactions have fully consistent decisions.

\subsection{Self-organizing search (SOS) policy}
\label{sec_sos}

In the self-organizing search (SOS), a searcher does not have any prior knowledge about the environment or the number of searchers and targets.
As in natural self-organization systems, such as a school of fish or a flock of birds, the objective is to equip searchers with the abilities that 
\begin{enumerate}[label=(\arabic*)]
\item a single searcher can perform an effective search by itself when there are no targets or searchers in its local view; 
\item a searcher has a tendency to follow other visible searchers so that a flock of searchers can be formed since the natural flocking behavior can increase the harvesting efficiency, which is especially true with a bigger group \cite{pitcher1982fish};
\item a flock of searchers can perform effective ``migration"–like actions rather than tangling with each other so that the flock as a whole loses searching ability.
\end{enumerate}
To achieve these goals, we use the actor-critic algorithm \cite{konda1999actor} to enable self-organizing searchers to learn from experiences in the centralized training and decentralized execution way.

The parameter $\theta$ of policy $\pi_\theta$ is updated with the learning rate $\alpha_1$ ($3\times10^{-4}$ and $10^{-4}$ in the search and pursuit experiments, respectively) according to
\begin{equation}\label{eq_rl_gradient}
\theta = \theta + \alpha_1 \bigtriangledown_\theta J(\pi_\theta),
\end{equation}
where
\begin{equation}
\bigtriangledown_\theta J(\pi_\theta) = E_{\tau  \sim \pi_\theta}[\sum_{t=0}^{tmax}{\bigtriangledown_\theta \log{\pi_\theta(a_t | s_t)} A_t]},
\end{equation}
and 
$\tau = (s_0, a_0, r_0, s_1, a_1, r_1, ...)$ is the trajectory;
$A_t$ is the generalized advantage estimation (GAE)  \cite{schulman2015high} in the form of
\begin{equation}\label{eq_rl_gae}
A_t = \sum_{l=0}^{tmax - t}(\gamma \lambda)^l \delta_{t+l}^V,
\end{equation}
with $\gamma, \lambda$ being two constants (0.99 and 0.97 in our experiments) and
\begin{equation}
\delta_{t}^V = R(s_t, \vec{a}_t) + \gamma V_{\phi}(s_{t+1}) - V_{\phi}(s_t).
\end{equation}
being the temporal difference (TD) residual of the approximate value function $V_{\phi}(\cdot)$ with discount $\gamma$.

The parameter $\phi$ of the value function $V_{\phi}(s_t)$ is optimized by minimizing the following loss function with stochastic gradient descent and learning rate $\alpha_2$ ($10^{-3}$ and $10^{-4}$ in the search and pursuit experiments, respectively):
\begin{equation}
\phi = \argmin_{\phi} E_{s_t, \hat{R}_t \sim \pi_{\theta}}[(V_{\phi}(s_t) - \hat{R_t})^2].
\end{equation}
where $\hat{R_t} = \sum_{t'=t}^{tmax} \gamma^{t' - t} R(s_{t'}, \vec{a}_{t'})$ is the discounted return from point $t$ with reward function $R(s_{t'}, \vec{a}_{t'})$ and discount factor $\gamma$.

\textit{\textbf{Reward function.}}
For the SOS task, individual agent's reward function $R^i(s_t, \vec{a_t})$ in the POMG is given in Table \ref{tbl_reward_function_sos}.
Though simple, experiments show that it achieves satisfied cooperation, and no additional efforts in the multi-agent credit assignment are needed as in the Dec-POMDP formulation.

\begin{table}[htbp]
	\begin{center}
	\caption{Reward function $R^i(s_t, \vec{a}_t)$ for self-organizing search (SOS)}	
	\label{tbl_reward_function_sos} 
	\begin{tabular}{l|l}
	\hline\hline
	Action & Reward \\
	\hline
	Search for a target & 10 to the contributing agent \\
	Collide with another agent & -12 $\times$ \# of agents collided with \\
	Collide with an obstacle & Die in its location \\
	Move before termination & -0.05 \\
	\hline
	\end{tabular}
	\end{center}
\end{table}

We once try to give the search reward to the contributing flock, which is a connected component of the graph whose vertexes are agents and edges represent local observations among agents.
We assume that if one member agent searches for a target, the whole flock of agents obtain the reward equally to encourage flocking behavior.
However, with such a reward mechanism, agents tangle with each other in local regions, although they indeed prefer gathering.
Instead, when we simply give a reward only to the contributing agent that finds the target, as in Table \ref{tbl_reward_function_sos}, the training performance improves significantly. 

Note that, the episode reward is defined as the mean of all agents' discounted accumulated rewards in the same environment.
In this way, the episode reward score will not increase with the number of agents involved, and thus, the scores are comparable between trials with different numbers of agents.

\textit{\textbf{Parameter sharing based centralized training.}} 
In training, agents in the same environment instance maintain a central experience pool and train shared critic and actor models with their newest collective episode experiences. 
The shared models in different environment instances are coordinated by communicating and averaging their gradients to stabilize the training.

\subsection{Cooperative coevolution algorithm for robots (CCR)}
\label{sec_ccr}

According to FSC2 (Algorithm \ref{algorithm_FSC2}), after distributed task allocation, the mission of a free pursuer is to cooperate with other cluster members pursuing the targeting cluster center.
For the single-target pursuit, we propose the CCR (cooperative coevolution for robots) algorithm based on CCPSO-R \cite{SUN2020CCPSOR,SUN2021CECBiQAP}, which further improves the cooperation of pursuers in their simultaneously decision making and execution process.

\textit{\textbf{Cooperative coevolutionary evaluation scheme.}}
Similar to CCPSO-R \cite{SUN2020CCPSOR}, the real agents in the CCR are the pursuers that execute physical actions in the environment, which can be represented by 2-D positions $\{A_i, i = 1,...,n \}$.
For each real agent $A_i$, all the neighboring positions one step away from it, including its current position, form a group of virtual agents $\{A_i^1 = A_i, ..., A_i^5\}$ that can act as the candidate next positions for the real agent.
The decision-making process of a real pursuer is to evaluate its virtual agents in the cooperative coevolutionary scheme and greedily select the best one as its next position.
The pursuit performance is ensured by the evaluation quality of the virtual agents, i.e., how well the fitness function is designed to guarantee conflict-free efficient cooperation in the pursuit.

In particular, the cooperative coevolutionary evaluation scheme means that the fitness evaluation of an individual agent is not only determined by itself, but also by the other real agents.
For the target cluster center $T_c$ and pursuer cluster $\{A_1,..., A_i^j, ..., A_{n^i}\}$, where the $i$-th member $A_i^j$ is the $j$-th virtual agent of the $i$-th real pursuer and $n_i$ is the total number of cluster members, the fitness function $f^{ij}_{stp}$ was proposed in CCPSO-R \cite{SUN2020CCPSOR} as follows:
\begin{equation}\label{eq_fit_stp}
f^{ij}_{stp} = f^{ij}_{closure} + f^{ij}_{expanse} + f^{ij}_{uniformity},
\end{equation}
where
\begin{equation}\label{eq_fit_closure}
f^{ij}_{closure} = inconv(T_c, A_1,..., A_i^j, ..., A_{n^i})
\end{equation}
evaluates whether the target $T_{c}$ is located in the convex hull formed by the pursuer cluster: 0 indicates that it is inside, 0.5 indicates that it is on the edge, and 1 indicates that it is outside;
\begin{equation}\label{eq_fit_expanse}
f^{ij}_{expanse} = \frac{1}{{n^i}}(\sum_{k=1,k \ne i}^{{n^i}}{||A_k - T_c||_1} + ||A_i^j - T_c||_1)
\end{equation}
gives the spatial extent of the pursuer cluster in terms of $T_c$; and
\begin{equation}\label{eq_fit_uniformity}
f^{ij}_{uniformity} = std
\left(\left[
\begin{array}{cc}
	N_{11} & N_{12}  \\
	N_{21} & N_{22}
\end{array}
\right]\right)
\end{equation}
or
\begin{equation}\label{eq_fit_uniformity_alternative}
\begin{array}{cc}
f^{ij}_{uniformity} = &std([N_{12},N_{21},N_{23},N_{32}]) \\
					         &+ std([N_{11},N_{13},N_{31},N_{33}]).
\end{array}
\end{equation}
evaluates how evenly the pursuer cluster is distributed around $T_c$ based on the standard deviation $std(\cdot)$ where $N_{ij}$ is the number of pursuers in the $(i, j)$-th space bin (for details, see \cite{SUN2020CCPSOR}).

However, $f^{ij}_{stp}$ only solves the cooperative single-target pursuit problem by letting agents make decisions sequentially, while its parallel decision-making version PCCPSO-R \cite{SUN2021CECBiQAP} can only resolve partial conflicts by introducing two secure distances in the fitness function.
Hence, we propose a new fitness function based on $f^{ij}_{stp}$ to enable conflict-free cooperation in single-target pursuit.
In detail, the fitness function for the $j$-th virtual agent of the $i$-the real pursuer $A_i^j$ can be defined as
\begin{equation}\label{eq_fitness}
f^{ij} = 
\begin{cases}
\infty, & \text{if } nnd^{ij}_{entity} == 0 \text{ or } \\
		& (nnd^{ij}_{target} \ne 1 \ \&\  nnd^{ij}_{pursuer} == 1) \\
f^{ij}_{convention}, & \text{else if } nnd^{ij}_{target} == 1 \ \&\ \\
					 & nnd^{ij}_{pursuer} == 1 \\
f^{ij}_{stp}, & \text{else}
\end{cases}
\end{equation}
where $nnd^{ij}_{entity}$ is the distance to the nearest neighbor with the set $entity$, which could be pursuers, targets or obstacles.
In the simultaneous decision-making and execution process, the secure distance between a pursuer and a target is 1 and that between pursuers is 2 to ensure that there are no collisions, and pursuers are not allowed to approach closer than this limit unless they are capturing a target.
However, when the condition $(nnd^{ij}_{target} == 1 \ \&\  nnd^{ij}_{pursuer} == 1)$ is satisfied, it means that  more than one pursuers may choose to occupy the same capturing position in the next step, where a conflict may occur but can be resolved by the lexicographic convention fitness function $f^{ij}_{convention}$ as follows.

\textit{\textbf{Lexicographic convention.}}
In the proposed lexicographic ordering, 2-D positions are sorted first in the ascending order of their first-dimension values and then based on their second-dimension values, and this is known by all agents.
This is used in the lexicographic convention that pursuers coordinate their choices of one-step-away open capturing positions by the following steps.
\begin{enumerate}[label=(\arabic*)]
\item All local open capturing positions are sorted.
\item All local free pursuers are sorted.
\item The neighboring open capturing positions and pursuers are paired in the priority order. 
\end{enumerate}
If the next candidate position or virtual agent $A_i^j$ of the current real pursuer $A_i$ is its assigned capturing position under a certain partial observation, $f_{convention}=-1$; otherwise, $f_{convention}=\infty$, which means that the choice not satisfying the lexicographic convention is not allowed.

\begin{figure}[htb!]
	\centering
	\subfloat[Incorrect decisions by {\color{blue}{A3}} due to its observation uncertainty.]{\makebox[1.5\width]{
	\includegraphics[width=0.3\linewidth]{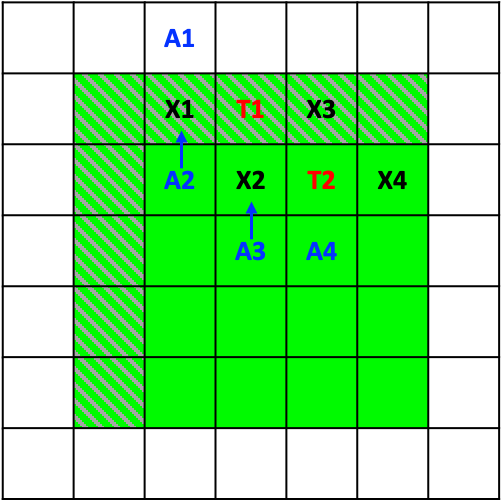}}
	\label{fig_risky_convention_1}}
	\quad
	\subfloat[Actual decisions of pursuers.]{\includegraphics[width=0.3\linewidth]{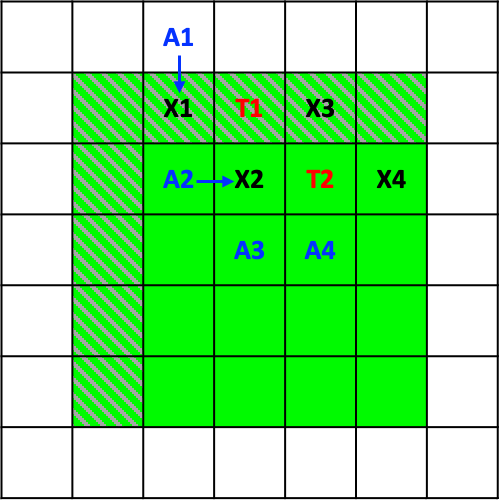}
	\label{fig_risky_convention_2}}
	\caption{Illustration of uncertain partial observation under the lexicographic convention of Section \ref{sec_ccr}; collisions may result if such scenarios are not detected. 
	{\color{blue}{A1, A2, A3, and A4}} are the pursuers, {\color{red}{T1 and T2}} are the targets, X1, X2, X3, and X4 are the open capturing positions, and these entities are numbered in the lexicographic order given in Section \ref{sec_ccr}. 
The {\color{green}{green}} background is the perception range of {\color{blue}{A3}}, and the dashed regions are the specific boundaries where risky capturing positions may appear. For {\color{blue}{A3}}, X1 is a risky capturing position that is located on the specific boundary of its local view and is assigned to a local free pursuer based on the local lexicographic convention without the detection of such scenarios. Meanwhile, the assigned capturing position X2 of {\color{blue}{A3}} has another neighboring free pursuer {\color{blue}{A2}}.
The decision of {\color{blue}{A3}}, which is made based on uncertain observation satisfying the above two conditions as in (a), may deviate from the actual decisions of pursuers as in (b) and risk collisions.}
	\label{fig_risky_convention}
\end{figure}

\textit{\textbf{Concept of certain partial observation.}}
The concept of certain partial observation is introduced to ensure multi-agent collision free in the pursuit.
It is in contrast to the uncertain partial observation, which is defined as the partial observation that satisfies the following two conditions, as illustrated in Figure \ref{fig_risky_convention}.
First, there exist risky capturing positions, which are the open capture positions on specific boundaries of the local view that will be assigned to a local free pursuer based on the lexicographic convention.
Second, there are other free pursuers neighboring the assigned captured position.
Under such uncertain observations, an agent may make risky decisions that may lead to collisions.
For simplicity, we prevent the current agent from taking the assigned capturing position by setting $f_{convention}=\infty$.
Although this may influence the efficiency, it can ensure that there are no collisions in the single-target pursuit due to the observation uncertainty in the POMG.

%

\section{Experiments}\label{sec_experiments}

\subsection{Environments, baselines, and experimental setups}

\paragraph{\textbf{Environments}}
First, for the convenience in comparing the self-organizing search (SOS) agents trained by  different MARL algorithms with their official public code, we made several changes to the PettingZoo Pursuit-V3 environment \cite{terry2020pettingzoo}, including the initialization, reward function, some utility functions, and bugs.
Second, for the multi-target self-organizing pursuit (SOP), we implemented the environment ourselves with more compact code and adjusted to the self-organizing game setups.
The local observation $o^i(s)$ of agent $i$ is always represented as an $11\times11\times3$ binary matrix, where the 3 channels are for targets, agents, and obstacles.
All code is available at https://github.com/LijunSun90/pursuitFSC2.

\paragraph{\textbf{Baselines for SOS}}
In the performance comparison of self-organizing search (SOS), we compare the actor-critic trained search policy with the following search strategies.
\begin{itemize}[leftmargin=*]
\item A swarm of independent random-walk searchers: Each searcher randomly walks in the space, taking no account of its surroundings and past history.
\item A swarm of independent complete searchers: 
A complete searcher searches the space in a systematic way to ensure that every position on the map is visited at least once.
This search is complete so that all targets are guaranteed to be found without a time limit.
The optimal systematic search strategy is a solution to the Hamiltonian path problem where every position is visited exactly once, which is NP-complete \cite{Garey1979Computers}.
For simplicity, we employ an intuitive systematic strategy in which the searcher first moves to its nearest map corner and then, starting from that corner, performs zigzag or snakelike walking assuming that the searcher knows the scope of the grid world but does not know the targets' positions.
Since the search success is defined as the agent occupying the target's position, the simple systematic searcher is actually equivalent to a searcher with a perception range of 1.
\item A swarm of ApeX-DQN searchers, the current documented best performing MARL in pursuit \cite{terry2020revisiting}:
We tested the learning rates \{$10^{-6}$, \boldsymbol{$10^{-5}$}, $10^{-4}$, $10^{-3}$\}; the batch sizes \{128, 256, 512, \textbf{1024}\}; the rollout fragment lengths \{\textbf{32}, 128\}; and Adam epsilons \{\textbf{0.00015}, $10^{-8}$\}, where the best values are shown in bold, and the other parameter values are the same as in \cite{terry2020revisiting}.
\item A swarm of coordinated MADDPG searchers: 
The OpenAI MADDPG implementation \footnote{https://github.com/openai/maddpg} is used in which an agent has access to all other agents' observations and actions through interagent communication; these are used in training the critic function $Q(\vec{o}, \vec{a})$.
We tested the learning rates \{$10^{-4}$, \bm{$10^{-3}$}, $10^{-2}$\}; the batch sizes \{256, \textbf{512}, 1024\}; and the model update rates \{4, \textbf{100}, 500\}, where the best values are shown in bold.
\end{itemize}

\paragraph{\textbf{Baselines for SOP}}
In the overall performance of the multi-target self-organizing pursuit (SOP), we compare tree implicit coordination methods: FSC2 and three others trained by the following MARL algorithms.
\begin{itemize}[leftmargin=*]
\item Actor-critic \cite{sutton2018reinforcement} (with parameter sharing): practical well-performed RL algorithm which is suitable for large-scale homogeneous agents. 
We tested the learning rates \{\bm{$10^{-4}$}, $5\times10^{-4}$\} (best value in bold) and three reward functions.
Besides, we trained the value function \{$10$, \bm{$80$}\} times every training epoch and got similar final performance.
The other hyperparameters are the same as those for the actor-critic algorithm in the self-organizing search experiments.
\item MAPPO \cite{yu2022surprising}: state-of-the-art on-policy MARL algorithm, which has the potential for large-scale applications.
We tested two inputs to the centralized value function: \{\textbf{concentration of all agents' observations}, agent-specific global state\} (similar), the learning rates \{\bm{$10^{-4}$}, $5\times10^{-4}$\} (best value in bold), three reward functions, and train both policy and value functions 10 times per training epoch.
The implementation is based on the official code \footnote{https://github.com/marlbenchmark/on-policy}, and other hyperparameters are the default values provided by \cite{yu2022surprising} for the MPE environments, which we verified with small experiments.
\item IPPO \cite{schulman2017proximal} (with parameter sharing): independent proximal policy optimization (PPO) algorithm with the same local observation as input for both the policy and value functions like the actor-critic baseline.
All the other hyperparameters are the same with MAPPO.
\end{itemize}

\paragraph{\textbf{Common experimental setup}}
The policy and value models in all MARL algorithms use the same architecture: two-layer ReLU multi-layer perceptions (MLP) with hidden layers of size 400 and 300.
In the SOP task, layer normalization \cite{ba2016layer} is added to each hidden and output layer for the three baseline algorithms: actor-critic, PPO, and MAPPO.
\begin{itemize}[leftmargin=*]
\item \textbf{Reward function:} 
For self-organizing search (SOS) tasks, all MARL algorithms use the same reward function in Table \ref{tbl_reward_function_sos}.
For self-organizing pursuit (SOP) tasks, all MARL algorithms use the same reward function in Table \ref{tbl_reward_function_sop}.
\end{itemize}

\begin{table}[htbp]
\begin{center}
\caption{Reward function $R^i(s_t, \vec{a}_t)$ for self-organizing pursuit (SOP)}	
\label{tbl_reward_function_sop} 
\begin{tabular}{l|l}
\hline\hline
Action & Reward \\
\hline
Capture a target & 10  \\
Neighbor a target & 0.1 \\
Collide & -12 \\ 
Move before termination & -0.05 \\
\hline
\end{tabular}
\end{center}
\end{table}

\begin{table*}[htbp!]
\begin{center}
\caption{Performance comparison on multi-target self-organizing pursuit (SOP) with 16 agents and 4 targets in $40\times40$ grid worlds. 
FSC2-HC: replace FSC2 fuzzy clustering with hard clustering. 
FSC2-NM: remove agent's memory in fuzzy clustering.
FSC2-RC: replace FSC2 fuzzy clustering with random clustering. 
FSC2-FS: replace FSC2 search with fish flocking rules (no migratory urge). 
* represents the statistical significance by student's $t$-test at the significance level 0.01.}	
\label{tbl_overall_performance_comparison} 
\begin{tabular}{c|c|c||c|c|c||c||c|c|c}
\hline\hline
\multicolumn{2}{c|}{Algorithm}
& FSC2
& FSC2-HC
& FSC2-NM
& FSC2-RC
& FSC2-FS 
& Actor-critic
& IPPO
& MAPPO
\\ \hline
\multirow{2}{*}{Clustering} 
& Memory
& \cmark
& \cmark
& \xmark
& \xmark
& \cmark
& -
& -
& -
\\ \cline{2-10}
{} 
& \tabincell{c}{Fuzzy\\membership} 
& \cmark
& \xmark
& \cmark
& \xmark
& \cmark
& -
& -
& -
\\ \hline
\multicolumn{2}{c|}{Self-organizing search}
& \cmark
& \cmark
& \cmark
& \cmark
& \xmark
& -
& -
& -
\\ \hline\hline
\multicolumn{2}{c|}{Capture rate} 
& \tabincell{c}{1\\(0)}
& \tabincell{c}{1\\(0)}
& \tabincell{c}{0.975*\\(0.075)}
& \tabincell{c}{0.965*\\(0.086)}
& \tabincell{c}{0.885*\\(0.155)}
& \tabincell{c}{0.91*\\(0.139)}
& \tabincell{c}{0.452*\\(0.228)}
& \tabincell{c}{0.6475*\\(0.223)}
\\ \hline
\multicolumn{2}{c|}{Episode length} 
& \tabincell{c}{108.09\\(50.256)}
& \tabincell{c}{119.35\\(75.99)}
& \tabincell{c}{156.39*\\(133.278)}
& \tabincell{c}{178.77*\\(147.369)}
& \tabincell{c}{308.94*\\(181.485)}
& \tabincell{c}{281.97*\\(176.28)}
& \tabincell{c}{496.06*\\(38.505)}
& \tabincell{c}{478.84*\\(63.689)}
\\ \hline
\multicolumn{2}{c|}{Collisions}  
& \tabincell{c}{0\\(0)}
& \tabincell{c}{0\\(0)}
& \tabincell{c}{0\\(0)}
& \tabincell{c}{0\\(0)}
& \tabincell{c}{0\\(0)}
& \tabincell{c}{6.59*\\(6.935)}
& \tabincell{c}{31.14*\\(25.285)}
& \tabincell{c}{12.47*\\(20.991)}
\\ \hline
\end{tabular}
\end{center}
\end{table*}

\begin{table*}[htbp!]
\begin{center}
\caption{Episode length (efficiency) comparison of FSC2 with fuzzy clustering and hard clustering on multi-target pursuit (SOP) in $40\times40$ grid worlds.
FSC2-HC: replace FSC2 fuzzy clustering with hard clustering.}	
\label{tbl_efficiency_of_fuzzy_vs_hard_clustering} 
\begin{tabular}{c|c|c|c|c|c|c|c}
\hline\hline
No. of agents 
& 16
& 32
& 64
& 128
& 256
& 512
& 1024
\\ \hline 
FSC2
& \tabincell{c}{\textbf{108.09}\\(\textbf{50.256})}
& \tabincell{c}{\textbf{114.51}\\(\textbf{69.885})}
& \tabincell{c}{134.22\\(102.551)}
& \tabincell{c}{\textbf{107.15}\\(\textbf{78.498})}
& \tabincell{c}{\textbf{112.83}\\(\textbf{119.471})}
& \tabincell{c}{\textbf{285.51}\\(\textbf{216.59})}
& \tabincell{c}{\textbf{338.18}\\(230.582)}
\\ \hline
FSC2-HC
& \tabincell{c}{119.35\\(75.99)}
& \tabincell{c}{115.09\\(83.229)}
& \tabincell{c}{\textbf{115.32}\\(\textbf{90.802})}
& \tabincell{c}{126.77\\(118.194)}
& \tabincell{c}{133.85\\(141.501)}
& \tabincell{c}{289.65\\(216.986)}
& \tabincell{c}{353.1\\(\textbf{224.407})}
\\ \hline
\end{tabular}
\end{center}
\end{table*}

\subsection{Self-organizing pursuit (SOP) experiments}\label{sec_exp_sop}

For the overall performance in multi-target self-organizing pursuit (SOP), we compare the proposed FSC2 method with three implicit coordination policies trained by the CTDE parameter sharing based actor-critic algorithm, PPO, and MAPPO, respectively.
These methods solve the large-scale implicit multi-agent coordination problem constrained by partial observation and no inter-agent communications in three ways: hierarchical decomposition, parameter sharing based coordinated reinforcement learning, and centralized value function enhanced coordinated reinforcement learning.

\begin{figure}[htb!]
	\centering
	\includegraphics[width=1.\linewidth]{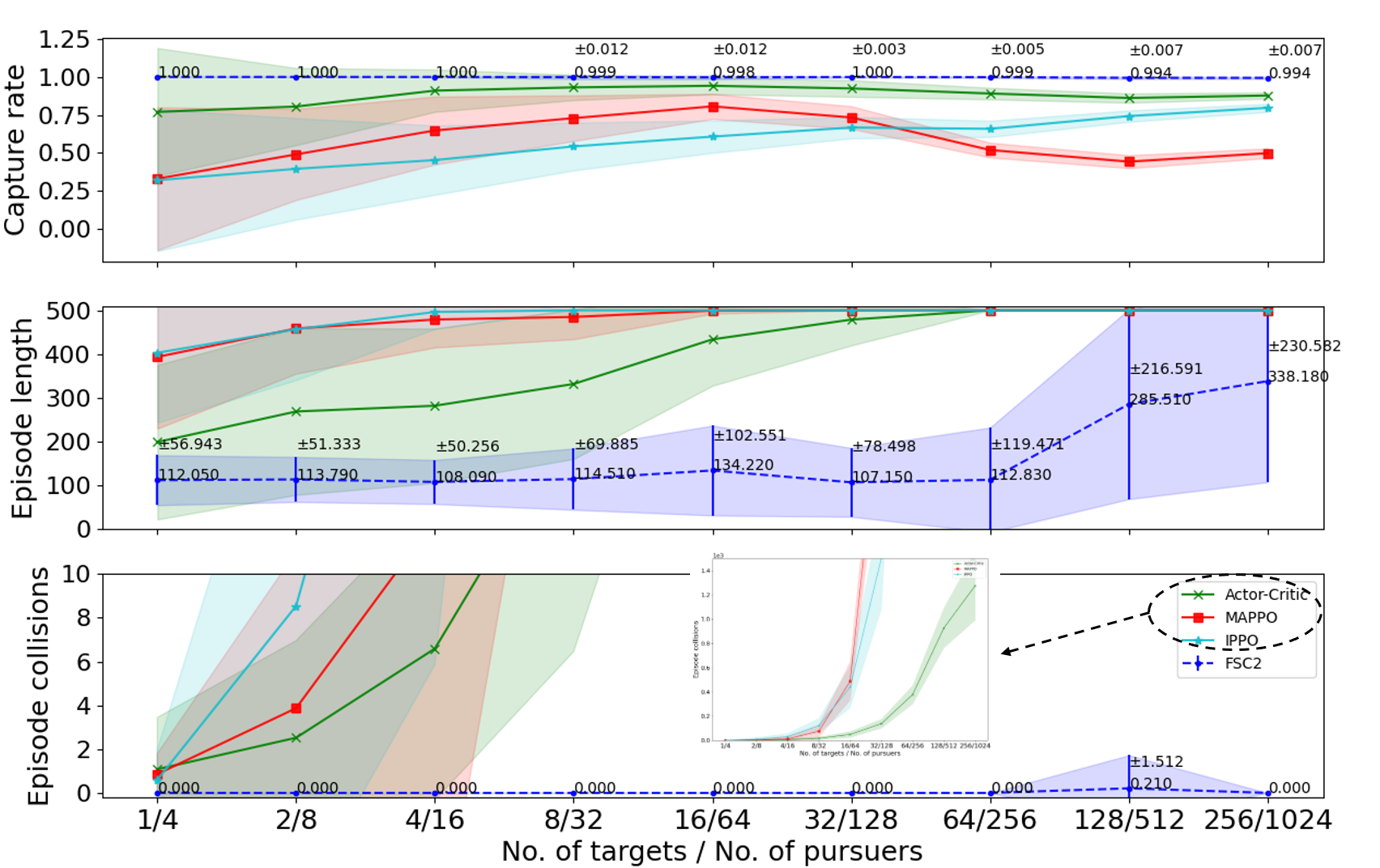}
	\caption{Swarm performance in the multi-target self-organizing pursuit (SOP) in $40\times40$ grid wolds with different numbers of targets and pursuers, where the mean and standard deviation of the experimental results in 100 independent runs are plotted.}
	\label{fig_sop_swarm_performance_diff_n_targets_40x40}
\end{figure}

The results are shown in Figure \ref{fig_sop_swarm_performance_diff_n_targets_40x40} and Table \ref{tbl_overall_performance_comparison}.
It can been seen that FSC2 significantly outperforms the other methods over all metrics.
Compared with PPO-based methods, the actor-critic algorithm achieves better results even with less model updates in the training.
Compared with IPPO, MAPPO performs better most of the time but its centralized value function does not achieve better multi-agent collision avoidance when the swarm density is extremely higher than that in its training.

From the video rendering results, the swarm search strategy, especially the swarm migration ability (see Section \ref{sec_sos}), plays an vital role in the overall performance, the ineffectiveness of which contributes to the inferior performances of general MARL policies.
Besides, another main challenge of general MARL algorithms is the multi-agent safety issue, such as the collisions.
It is very hard to achieve the safety guarantee by a reward function, which is especially challenging with more agents and conflicts of interests being involved \cite{zhang2021multi}. 
The conflicts are non-trivial to be resolved since agents make decisions and execute actions simultaneously in POMG.
In contrast, FSC2 employs the CCR algorithm as the third module in its framework for the close coordination of agents, the safety of which is guaranteed by the fitness function in the online planning.

\begin{figure}[htb!]
	\centering
	\includegraphics[width=1.\linewidth]{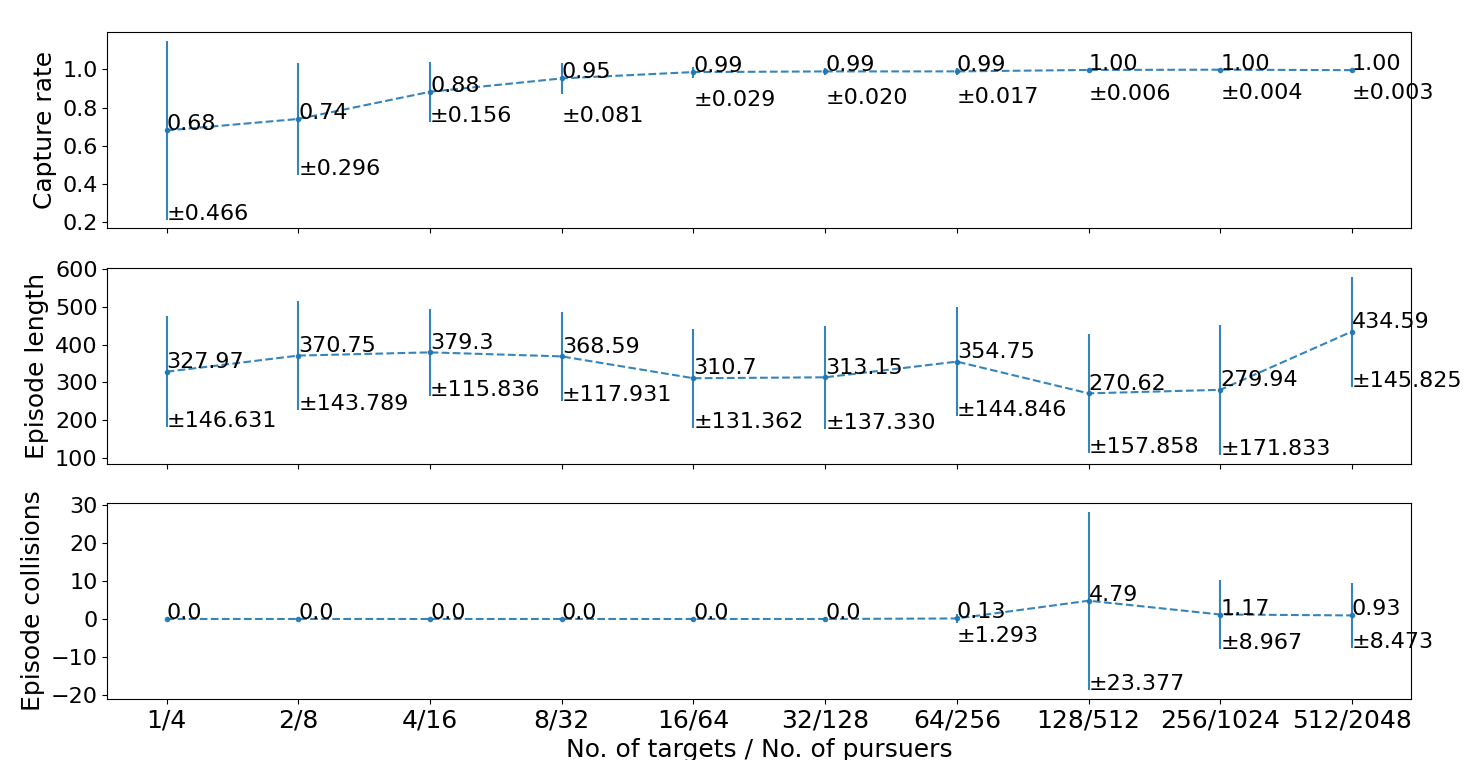}
	\caption{Swarm performance in the multi-target self-organizing pursuit (SOP) in $80\times80$ grid wolds with different numbers of targets and pursuers, where the mean and standard deviation of the experimental results in 100 independent runs are plotted.}
	\label{fig_sop_swarm_performance_diff_n_targets_80x80}
\end{figure}

\paragraph{\textbf{Scalability and swarm performance of FSC2}}
We test the swarm performance and scalability of up to 2048 FSC2 agents in multi-target SOP in $40\times40$ and $80\times80$ grid worlds, as shown in Figures \ref{fig_sop_swarm_performance_diff_n_targets_40x40} and \ref{fig_sop_swarm_performance_diff_n_targets_80x80}, respectively.
Almost all experiments achieve a nearly 100\% average capture rate except that when the number of pursuers is too small to cover the space in the maximum of 500 time steps, such as in the cases of 4 and 8 pursuers in $80\times80$ grid worlds in Figure \ref{fig_sop_swarm_performance_diff_n_targets_80x80}.
However, the more than 68\% average capture rate proves the efficient search ability of FSC2 agents in such trials.
\qquad
\begin{figure}[htb!]
	\centering
	\includegraphics[width=0.8\linewidth]{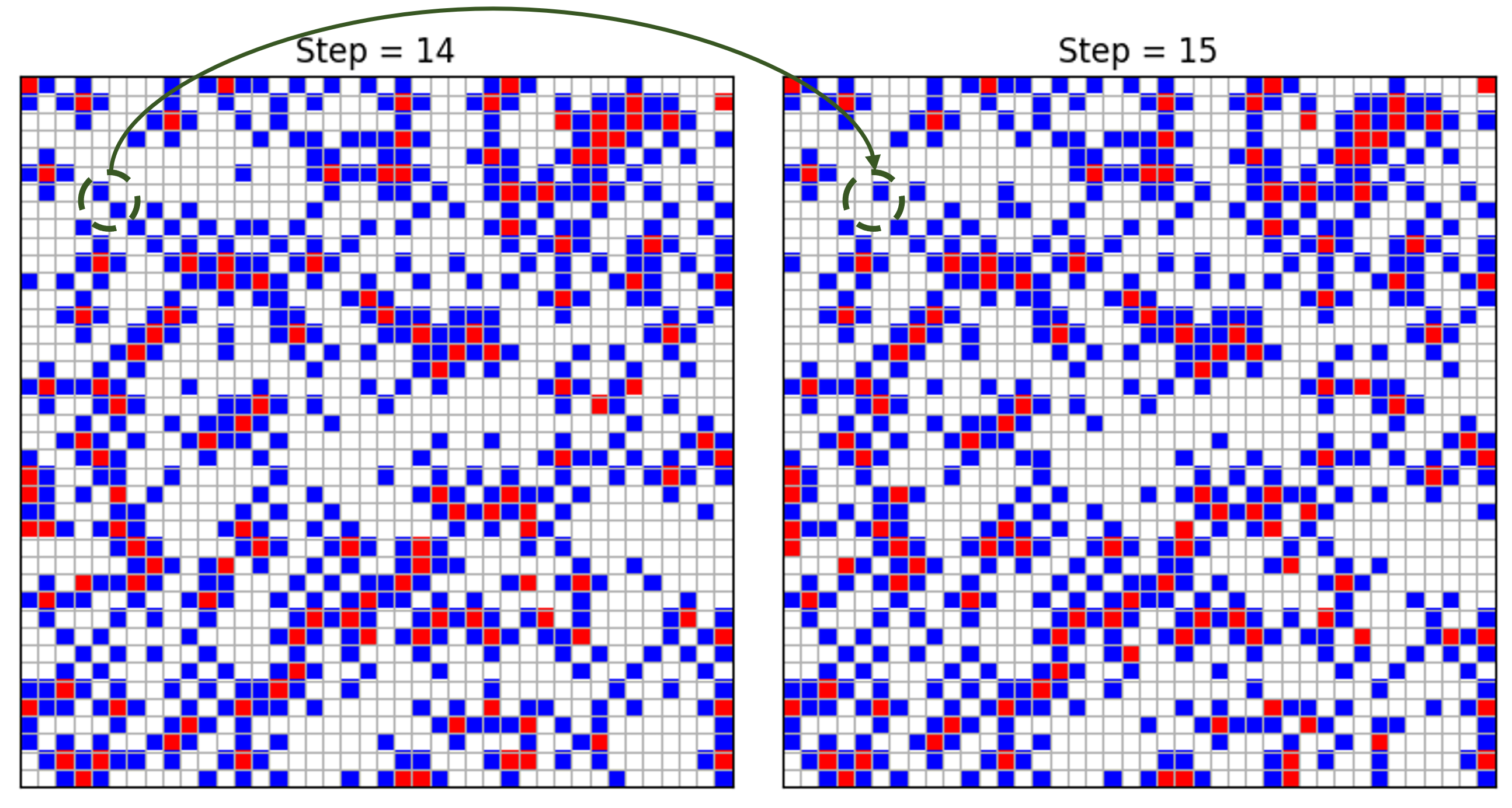}
	\caption{Multi-agent collision scenario illustration in the multi-target SOP with 128 targets ({\color{red}{red squares}}) and 512 agents ({\color{blue}{blue squares}}) in $40\times40$ grid world: the two circled agents in step 14 are two searchers that collide with each other in step 15.}
	\label{fig_sop_collision}
\end{figure}

Note that, the collisions in 0.22\% and 2.1\% of the trials in Figures \ref{fig_sop_swarm_performance_diff_n_targets_40x40} and \ref{fig_sop_swarm_performance_diff_n_targets_80x80} occur when the FSC2 agent is a searcher, i.e., the SOS agent in Algorithm \ref{algorithm_FSC2}.
This does not mean a performance degradation of SOS agents in SOP tasks.
Rather, it reveals the weak safety guarantee of RL algorithms.
Figure \ref{fig_sop_collision} gives two consecutive frames showing an inter-agent collision when 128 targets and 512 pursuers are deployed in the $40\times40$ grid world.
Although SOS agents learn to interact with each other in the multi-agent environment and the collisions are reduced significantly, it cannot be avoided absolutely.
In the search subtask, SOS agents are only trained in very simple environments where boundary walls are the only obstacles.
By deploying SOS agents in the multi-target SOP, however, they are often surrounded by increasingly complex distribution of captured targets and locked pursuers that are equivalent to obstacles, and the environment is more like a complicated maze.
Besides, compared with the collision avoidance with static obstacles, the multi-agent collision avoidance is a more complicated coordination problem that is harder to be fully guaranteed by RL.
In such scenarios, FSC2 agents can still capture nearly 100\% of the targets within the limit of 500 time steps without collisions most of the time, which can also be seen from the large standard deviation of the nonzero mean collisions in Figures \ref{fig_sop_swarm_performance_diff_n_targets_40x40} and \ref{fig_sop_swarm_performance_diff_n_targets_80x80}.

In addition, the relatively stable swarm performance of FSC2 agents indicates that the three proposed subsolutions in FSC2, i.e., the MARL-trained self-organizing search (SOS) agents, fuzzy-based distributed task allocation, and the CCR-based single-target pursuit, all fulfill their responsibilities effectively and efficiently, which also indicates the good scalability of FSC2 agents.
Due to the fully distributed nature of the proposed self-organizing algorithm FSC2, its application and performance are not restricted by the swarm size.

\subsection{Self-organizing search (SOS) experiments}\label{sec_exp_sos}

\begin{figure*}[htb!]
\centering
\subfloat{
  \centering
  \includegraphics[width=.19\linewidth]{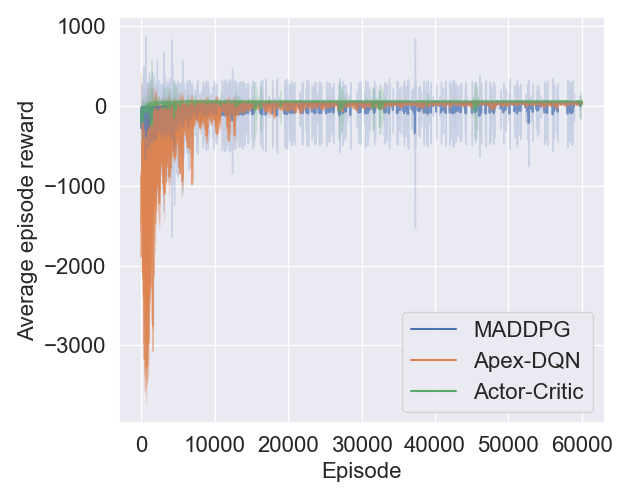}
  \label{fig_episode_reward}}
\subfloat{
  \centering
  \includegraphics[width=.19\linewidth]{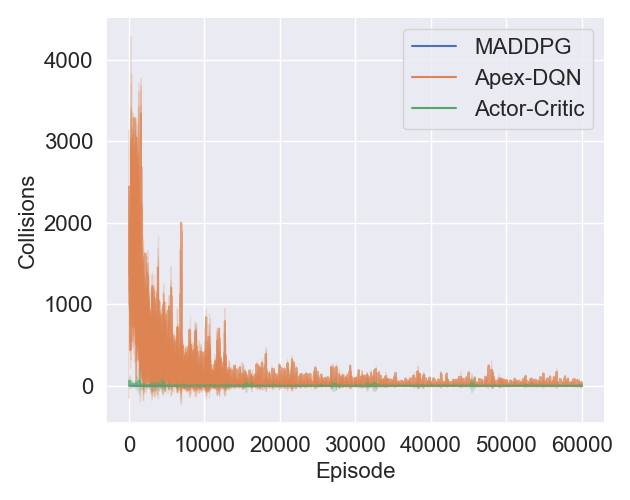}
  \label{fig_episode_collisions}}
\subfloat{
  \centering
  \includegraphics[width=.19\linewidth]{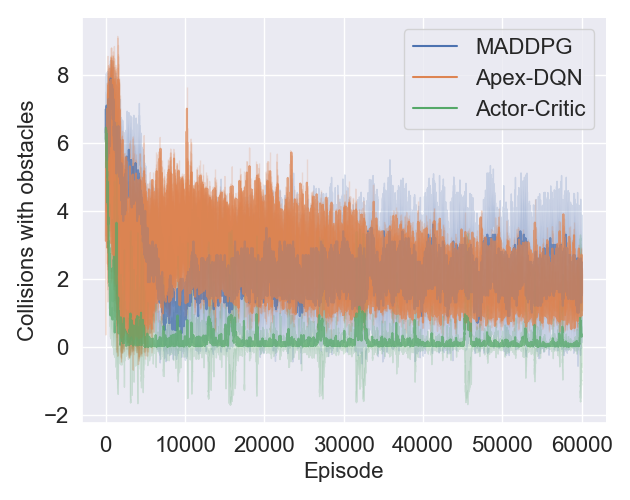}
  \label{fig_episode_collisions_with_obstacles}}%
\subfloat{
  \centering
  \includegraphics[width=.19\linewidth]{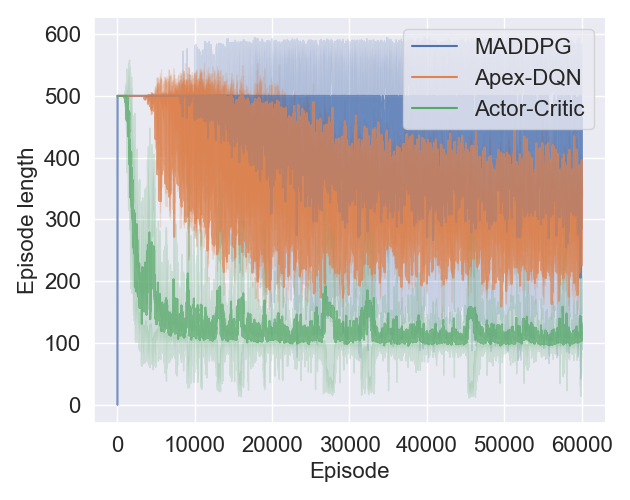}
  \label{fig_episode_length}}%
\subfloat{
  \centering
  \includegraphics[width=.19\linewidth]{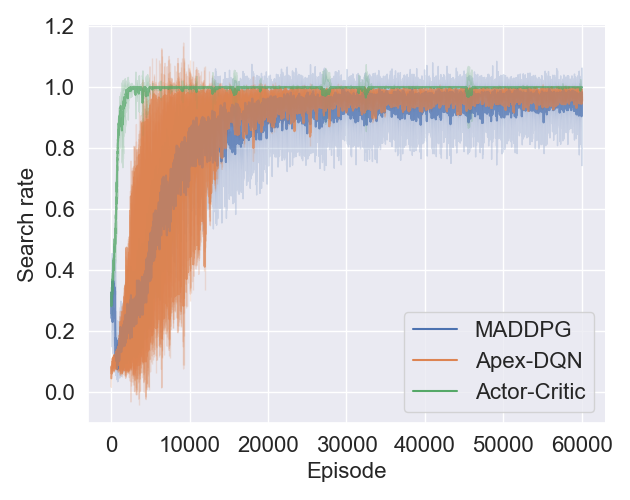}
  \label{fig_episode_capture_rate}}
\caption{Training performance comparison on self-organizing search (SOS) over 10 random seeds, where the solid lines and shaded areas represent the mean and standard deviation of the corresponding performance, respectively.}
\label{fig_training_performance}
\end{figure*}

The training performances of the actor-critic, ApeX-DQN, and MADDPG models for 8 agents searching 50 targets in $40\times40$ grid worlds are shown in Figure \ref{fig_training_performance}.
The average episode reward, episode length, number of collisions between agents, and number of collisions with obstacles all contribute to the reward received by agents as given in Table \ref{tbl_reward_function_sos} and thus the agents' training, while the episode search rate is not part of the reward function and is presented to illustrate the effectiveness of the training.

The actor-critic model has the best training performance in terms of convergence speed, the final converged values, and the stability of the training performance.
In contrast, both MADDPG and ApeX-DQN are influenced more by the random seeds in the training.
MADDPG oscillates severely during the training process.
Regarding to ApeX-DQN, we observed that the convergence speed is not the most important metric since its performance may degrade and diverge badly with a faster convergence speed.
Therefore, we chose the parameters that enable ApeX-DQN's performance to improve steadily, the final performance of which is proven to be better than the best training performance of the parameters with faster convergence that later degrade.

\begin{figure}[htb!]
	\centering
	\includegraphics[width=0.8\linewidth]{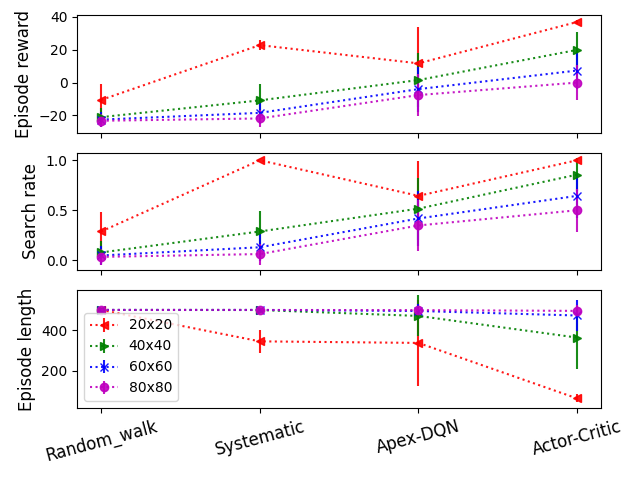}
	\caption{Single SOS agent performance comparison in grid worlds of different sizes, where the mean and standard deviation of the experimental results in 100 independent runs are plotted}
	\label{fig_single_searcher}
\end{figure}

Second, we compare a single agent's searching performance in the $20\times20, 40\times40, 60\times60$, and $80\times80$ grid worlds with 5 targets in Figure \ref{fig_single_searcher}.
With the increase of the environment size and the sparsity of targets, the performances of all policies change accordingly, and the actor-critic searcher is always the best.
For the random-walk searcher, the environment size has little influence on its performance due to its local random movements, which take longer to explore farther regions.
For the systematic searcher, when the environment size is too large to allow it to perform a complete systematic search in a limited time, its performance is slightly better than that of the random-walk searcher.
Therefore, compared with a complete searcher, the actor-critic searcher has better performance in searching targets in a limited time in most scenarios.

\begin{figure}[htb!]
	\centering
	\includegraphics[width=0.8\linewidth]{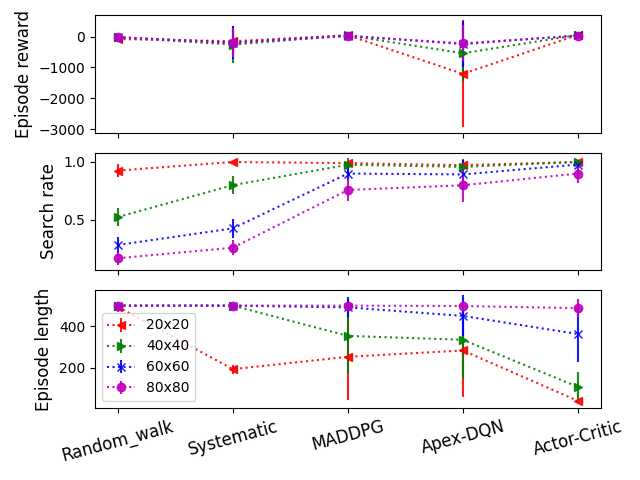}
	\caption{Swarm performance comparison of 8 SOS agents searching 50 targets in grid worlds of different sizes, where the mean and standard deviation of the experimental results in 100 independent runs are plotted}
	\label{fig_swarm_performance_same_swarm_size}
\end{figure}

Third, we compare the swarm performance of different policies by searching 50 targets with 8 searchers in $20\times20, 40\times40, 60\times60$, and $80\times80$ grid worlds, as shown in Figure \ref{fig_swarm_performance_same_swarm_size}.
The smaller the environment is, the larger the swarm density is, and the more challenging the mulit-agent coordination is; and the actor-critic swarm always performs best.
Although MADDPG is the algorithm that considers the multi-agent interactions the most in its critic function learning, its performance is not as good as that of actor-critic.
In addition, since MADDPG learns a unique critic function for each agent, when the number of agents changes, it needs to relearn.

Finally, the comparison of Figures \ref{fig_single_searcher} and \ref{fig_swarm_performance_same_swarm_size} proves two facts.
First, the superiority of a swarm of independent agents over a single-agent system stems from the benefits of introducing more agents, such as random-walk agents and systematic agents.
Second, coordinated inferior agents may sometimes outperform single superior agents in some aspects, such as the swarm of ApeX-DQN agents that outperform the single actor-critic agent.

\paragraph{\textbf{Explainable search behavior analysis and sparse targets exploration}}
One basic problem to be solved in self-organizing search is how a searcher behaves when there is no information (no targets and no other searchers) in its current perception, i.e., in the case of an empty observation.
To simulate natural flocking, Reynolds \cite{reynolds1987flocks} proposed three behavioral rules for individual agents: (1) avoid collisions with neighbors; (2) match velocity with neighbors, and (3) stay close to neighbors,
which also appear in the three behavior patterns of individual fish models in the movement of a school \cite{pitcher1982fish}.
However, as indicated in \cite{reynolds1987flocks}, these three behaviors can only support aimless flocking; it is also observed in our experiments that if we only apply these three rules, agents can group together yet become tangled with each other in local regions so that the whole group loses the search ability.

\begin{figure}[htb!]
	\centering
	\includegraphics[width=0.6\linewidth]{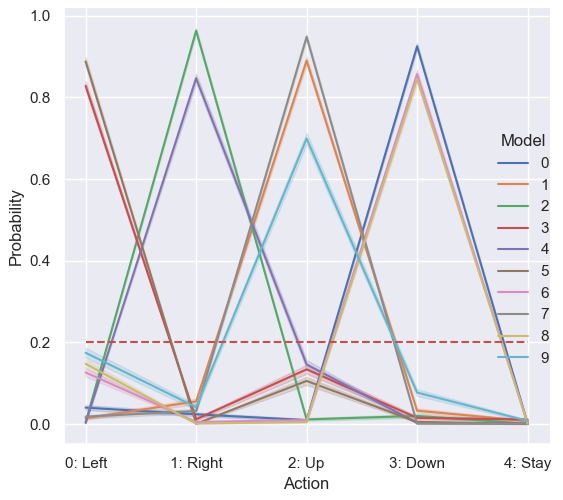}
	\caption{Behavior probability or action distribution of actor-critic trained self-organizing search (SOS) policy with the empty observation, which is estimated from 100 independent runs with 1000 steps per run. The different models are actor-critic policies trained with different random seeds.}
	\label{fig_action_distribution}
\end{figure}

Similar to the case of adding a global direction or global target as the flock's migratory urge in \cite{reynolds1987flocks}, we observe that the successfully trained self-organizing searchers learn similar behaviors by themselves. 
As shown in Figure \ref{fig_action_distribution}, we test the actor-critic searcher's behavior by always feeding it with the empty observation, and then estimate the searcher's action distribution over its 5 legal actions by running these tests in 100 independent runs with 1000 steps per run.

It can be seen that although different policies trained with different random seeds have different preferences, the common result is that they prefer a particular action most of the time and stochastically choose other actions.
In contrast to the random walk with a uniform action distribution, shown as the red dashed line in Figure \ref{fig_action_distribution}, this trained action distribution ensures that a searcher will move in one direction most of the time and occasionally switch to another direction, which benefits the target search since the searchers are moving farther away, exploring nonrepeatable areas most of the time, and covering a wide expanse of the map in a limited time.
This searching behavior also provides a way to the space exploration problem with sparse targets, as the example shown in Figure \ref{fig_single_searcher}.

In addition, since the self-organizing searchers are homogeneous, when all searchers perform similar behaviors, as a whole, the self-organizing search swarm behaves as an emergent self-organized pattern.
In other words, the self-organized pattern in the self-organizing search emerges here because the agents are homogeneous and behave according to the same meaningful actions.

\subsection{Consistency analysis in distributed task allocation}\label{sec_exp_consistency_analysis}

\begin{figure*}[htb!]
	\centering
	\subfloat[Illustrative scenarios: fuzzy clustering is stochastically superior to hard clustering. Note that, in $M^2$, the membership value of $A_1$ to $T_2$ is 0 because $A_2$ can infer that $T_2$ is outside the perception scope of $A_1$ as all agents are homogeneous and have the same perception radius.]{\includegraphics[width=.7\linewidth]{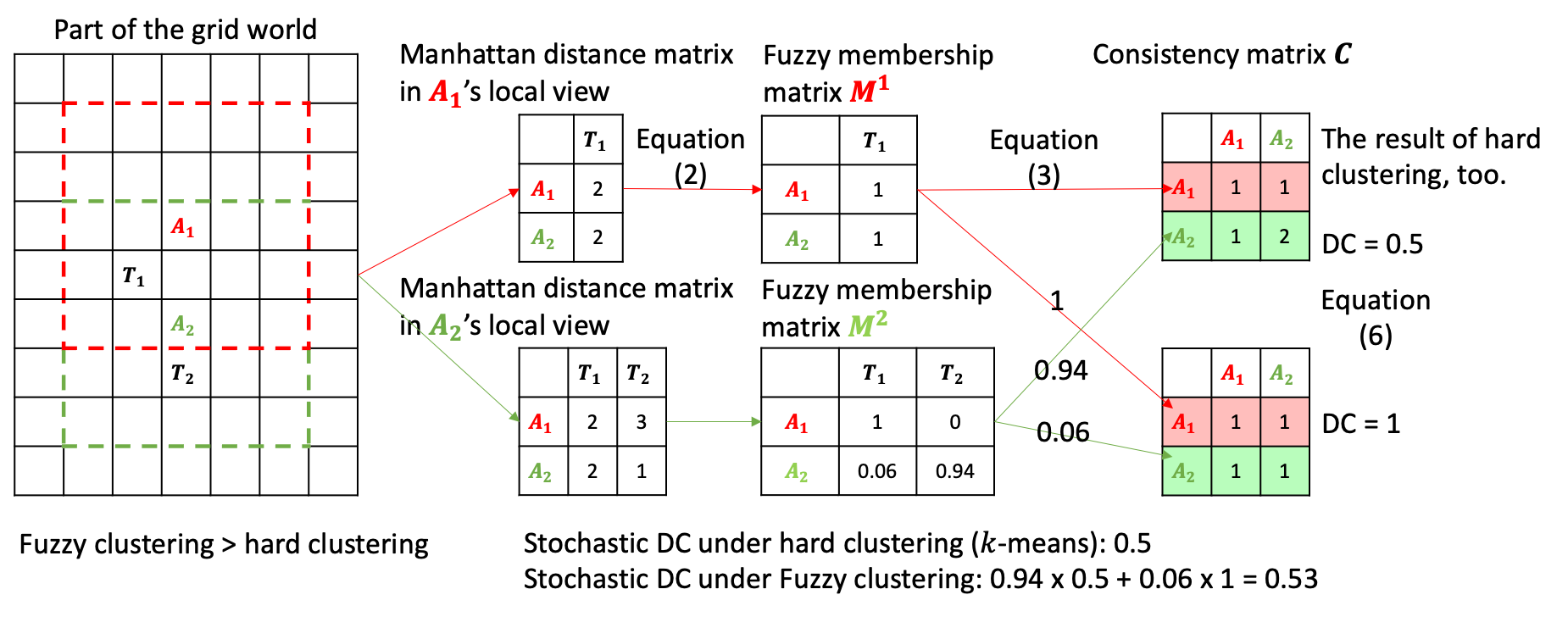}
	\label{fig_DC_fuzzy_superior}}
	\\
	\subfloat[Illustrative scenarios: fuzzy clustering is stochastically inferior to hard clustering.]{\includegraphics[width=.7\linewidth]{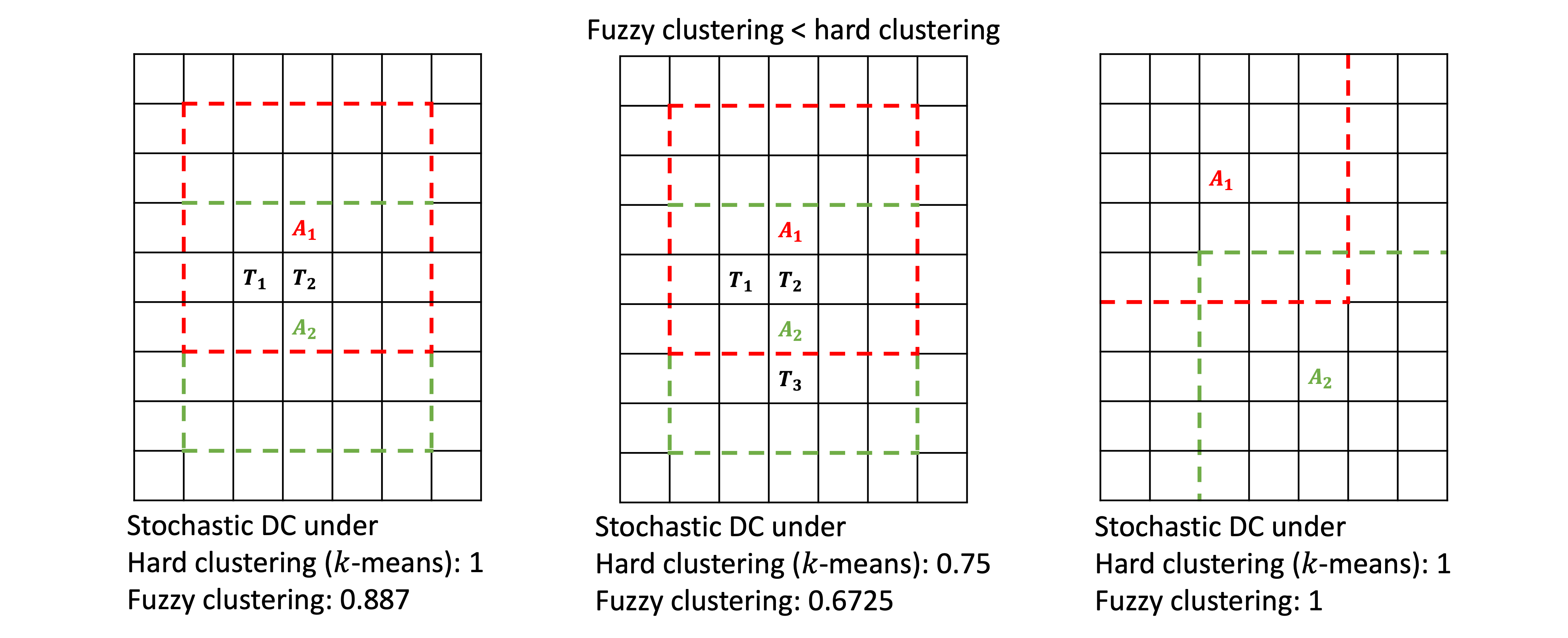}
	\label{fig_DC_fuzzy_inferior}}
	\caption{The computational process of DC in Equation (\ref{eq_dc}) and stochastic comparisons between fuzzy clustering and hard clustering in distributed task allocation, where the symbols ``$>$" and ``$<$" represent stochastically superior and inferior, respectively, and a dashed rectangle around an agent of the same color indicates its local perception scope with an $inf$-norm radius of 2 for the purpose of illustration.}
	\label{fig_analyze_fuzzy_clustering}
\end{figure*}

In the distributed task allocation, pursuers and targets are grouped into clusters such that the multi-target SOP is locally decomposed into several single-target pursuit problems.
However, in this distributed decision-making process, there may be inconsistency to some extent.
As illustrated in Figure \ref{fig_DC_fuzzy_superior}, due to the partial observability of pursuers, it is common that an agent can only observe part of another agent's local perception so that they have different knowledge of the world, which is the source of inconsistency in distributed clustering.

For hard clustering, such as $k$-means, an agent randomly selects one of its nearest targets as its cluster center, while for fuzzy clustering, the choice of targets is determined stochastically by the fuzzy membership matrices.
The random choices between the nearest targets in hard clustering and fuzzy membership values in the fuzzy clustering may all stochastically result in different consistency matrices $C$.
We multiply the $DC$ value of each matrix $C$ with its corresponding probability and obtain the stochastic $DC$ value.
Figure \ref{fig_DC_fuzzy_superior} gives an example scenario in which  fuzzy clustering is stochastically superior to hard clustering.
Such scenarios occur when the uncertainty outside of the common observation area brings better options for the agents, such as $T_2$ to $A_2$ in Figure \ref{fig_DC_fuzzy_superior}.
In contrast, as illustrated in Figure \ref{fig_DC_fuzzy_inferior}, fuzzy clustering is stochastically inferior to hard clustering when the uncertainty outside of the common observation area fails to provide better options for the agents, such as $T_3$ to $A_2$, and when there is no any uncertainty.

However, since uncertainty is inherent in the partially observable game, an agent can never determine the level of uncertainty from only its own local view without other related information communicated between neighboring agents.  
In addition, what is important here is that with fuzzy clustering, in scenarios where fuzzy clustering is stochastically inferior to hard clustering, its stochastic process enables it to be as good as or even better than hard clustering.
In contrast, with hard clustering, in scenarios where hard clustering is not stochastically superior to fuzzy clustering, its clustering result will never beat the fuzzy clustering result.
Therefore, fuzzy clustering reduces the influence of uncertainty in distributed task allocation in partially observable environments, especially in cases without interagent communication.

\subsection{Ablation studies}

\paragraph{\textbf{Influence of fuzziness: fuzzy clustering vs. hard clustering}}
We replace the fuzzy membership value calculation of Equation (\ref{eq_mu_ij}) to (\ref{eq_cluster_center}) in the fuzzy clustering with a hard clustering method.
Since the cluster centers are known to be local free targets as introduced in Section \ref{sec_fuzzy_clustering}, there is no need to calculate the $k$ cluster centers as $k$-means.
But, similar to the hard-clustering in $k$-means, an agent greedily selects the nearest cluster center and joins that cluster.
The result is shown in the column of FSC2-HC of Table \ref{tbl_overall_performance_comparison}.
In the 100 experiments, we observe the efficiency degradation of FSC2-HC in terms of the episode length, but we do not obtain the statistical significance evidence from student $t$-test.
Following the same statistical comparison of Table \ref{tbl_overall_performance_comparison}, we compare the efficiency of FSC2 and FSC2-HC in $40 \times 40$ grid worlds with the 16, 32, 64, 128, 256, 512, and 1024 agents and get the same conclusion, as shown in Table \ref{tbl_efficiency_of_fuzzy_vs_hard_clustering}.
The conclusion here is in accordance with the clustering consistency analysis that the fuzziness of fuzzy clustering and its stochastic clustering enable the fuzzy clustering to be as good as or even better than hard clustering.
In other words, the task time is extended due to the inconsistent distributed hard clustering.
Besides, another reason that we do not achieve the stochastic significance may be that we do not optimize the fuzzifier parameter $\alpha$ in Equation (\ref{eq_mu_ij}) and our test experiments are not large enough to observe the difference.

\paragraph{\textbf{Influence of memory}}
We remove the agent memory in the fuzzy clustering of Algorithm \ref{algorithm_fuzzy_clustering}.
The comparison result is shown in the FSC2-NM column of Table \ref{tbl_overall_performance_comparison}.
It is seen that without the agent memory, the multi-target pursuit performance degrade significantly.
As introduced in Section \ref{sec_fuzzy_clustering}, without the memory, an agent is hard to cope with temporal uncertainty due to the partial observation and may switch the roles between searcher and pursuer.
It causes the inconsistent or unstable successive decision-making of agents in the time scale and thus reduce the overall task performance.

\paragraph{\textbf{Influence of clustering}}
In this part, we totally replace the first module of FSC2: fuzzy clustering, with the random clustering to see to what extent an effective clustering method can influence the overall multi-target pursuit.
The result is shown in the FSC2-RC of Table \ref{tbl_overall_performance_comparison}.
It can be seen that, a random clustering performs significantly worse than FSC2, FSC2-HC, and FSC2-NM, which prove the necessity of effective clustering to the whole task completion.

\paragraph{\textbf{Influence of ``migration" ability in search}}
In Section \ref{sec_sos}, we propose three abilities for a successful self-organizing searcher: (1) the ability to effective search as a single agent; (2) the ability to form flocks and get benefits from the swarm;  and (3) the ability to perform effective ``migration"-like actions in order to realize the swarm potential.
In this part, we remove the third ability by replacing the second module of FSC2: RL trained search policy, with the three behavioral rules proposed by Reynolds \cite{reynolds1987flocks} in simulating natural flocking and school of fish \cite{pitcher1982fish}.
The three rules are: (1) avoid collisions with neighbors; (2) match velocity with neighbors, and (3) stay close to neighbors.
As indicated in \cite{reynolds1987flocks}, these three behaviors can only support aimless flocking, i.e., no ``migration" ability.
The result is shown in the FSC2-FS column of Table \ref{tbl_overall_performance_comparison}, which significantly perform worse than others in terms of both capture rate and efficiency.
It proves the importance of ``migration" ability in the self-organizing search.

\subsection{Discussion}\label{sec_discussion}

\paragraph{\textbf{Computational complexity analysis.}}
For a distributed partially observable agent without communication, the computational complexity is not related to the swarm size but only related to the observation range.
Assume that there are $n$ pursuers and $m$ targets in the local observation defined by the range $r$, where $n + m \le r^2$, and let $c_i, i = 1, 2, ...$ be some constants.
First, for the distributed task allocation in Section \ref{sec_fuzzy_clustering}, the time complexity in terms of Equations (\ref{eq_fuzzy_membership_matrix}) to (\ref{eq_cluster_members}) is $(c_1 \cdot n \cdot m + c_3 \cdot m) + c_3 \cdot n \cdot m + c_4 \cdot m + c_5 \cdot n = O(n \cdot m)$.
Second, for the SOS in Section \ref{sec_sos}, the time complexity of the policy model with input size $3r^2$ is $O(r^2)$.
Third, for the single-target pursuit in Section \ref{sec_ccr}, the time complexity \footnote{http://www.qhull.org/html/qh-code.htm\#performance} of Equation (\ref{eq_fit_stp}) is $O(nlogn) + c_1 \cdot n + c_2 \cdot (m + n) = O(nlogn)$, 
while the time complexity of calculating the lexicographic convention in Equation (\ref{eq_fitness}) is $O(n^2) + O(m^2) + O(n \cdot m) = O(max(n, m)^2)$ in the worst case.
Therefore, based on Algorithm \ref{algorithm_FSC2}, FSC2's time complexity is $O(max(n, m, r)^2)$ in the worst case.

\paragraph{\textbf{Generalization of FSC2 and comparison with existing work.}}
As introduced in Section \ref{sec_intro}, there are many capture definitions in the pursuit domain.
The proposed FSC2 algorithm can be extended to other multi-agent pursuit games, although it is originally proposed for the 4-pursuer-surrounding-based capture.
For example, FSC2 satisfies the mass capture based pursuit in \cite{ZHOU2021}.
In FSC2, when pursuers surround the target, the mass center of pursuers will match that of the target.
But instead of the mass center of the group including all pursuers matching that of the evader group and thus one mass capture in \cite{ZHOU2021}, four pursuers take charge of each target and thus there are many distributed mass captures in the FSC2.
Therefore, compared with the mean field reinforcement learning of Zhou et al. \cite{ZHOU2021}, FSC2 is more suitable for the pursuit where pursuers and targets are spatially distributed.
In particular, FSC2 can additionally deal with the interagent collision avoidance.
On the other hand, FSC2 can directly solve the pursuit problems with one more time step if the capture is occupying-based and the number of pursuers needed for a target is not greater than 4, as in MPE \cite{lowe2017multi}.
FSC2 agents only need to walk towards the target one more step after they surround the target and the target cannot move.
Actually, in addition to the occupying-based pursuit, pursuers can do many things as long as the target is surrounded, such as tagging the target as in MAgent \cite{Zheng2018MAgent}.
In the proposed fuzzy-based distributed task allocation, we do not limit the number of agents in a cluster to greedily capture one visible target with as many pursuers as possible.
This is beneficial when applying the FSC2 in other pursuit problems under the occupying-based capture yet with more than 4 pursuers for each target.
In addition, the fitness function, i.e., Equation (\ref{eq_fitness}), of the CCR algorithm is originally designed to suit the capture with more than 4 pursuers, as shown in its sequential decision-making version: CCPSO-R \cite{SUN2020CCPSOR}.
The only necessary modifications are the capture definition and the order of agents in which they walk toward the target to ensure that there are no collisions.

%

\section{Conclusion}\label{sec_conclusion}

This paper investigated the large-scale partial observable multi-target SOP problem by formulating it as a POMG and proposed the distributed algorithm FSC2 based on the fuzzy logic, MARL, and evolutionary computation.
It does not rely on interagent communication and is thus naturally robust to unavoidable communication failures in general multi-agent game setups.
In particular, FSC2 dealt with two kinds of uncertainties in SOP: observation uncertainty and interaction uncertainty. 
By comparing with other implicit coordination policies, we proved the superior performance of FSC2 and the benefits of the hierarchical framework by decomposing the task.
The scalability, interpretability, and rationality of FSC2 have been verified through experiments, empirical analyses, and ablation studies.


However, the safety of interagent collision avoidance is difficult to be guaranteed by MARL without explicit communications, which has also been verified by our experiments.
This was one motivation that we apply MARL only in the search sub-task, not the target pursuit task which needs more close coordination and challenges the RL methods more.
In future work, more complex self-organizing patterns are expected to emerge that are not simply due to homogeneous agents, and the distributed implicit multi-agent coordination problem needs to be further investigated, especially in terms of the multi-agent safety issue.

%
%


\bibliographystyle{elsarticle-num} 
\bibliography{mybibfile}

\end{document}